\documentclass[%
 reprint,
 amsmath,amssymb,
 aps,
prb,
]{revtex4-2}

\usepackage[utf8]{inputenc}
\usepackage{graphicx}
\usepackage{overpic}
 \usepackage{rotating}
 \usepackage{mwe, tikz}
 \usepackage{amsmath,amssymb}
\usepackage{pgfplots}
\usetikzlibrary{calc,arrows,decorations.markings,shapes.geometric} 
\usepackage{amssymb}

\usepackage[finalnew]{trackchanges}

\usepackage{lipsum}
\usepackage{romannum}
\usepackage{xfp}

\soulregister\ref7
\soulregister\cite7
\soulregister\citeA7
\soulregister\romannum7

\usepackage{graphicx}
\usepackage{amsmath}
\usepackage{mathtools} 
\usepackage{color}
\usepackage{overpic}
\usepackage[export]{adjustbox}
\usepackage{mwe, tikz}
\usepackage{rotating}
\usepackage{latexsym}
\usepackage{csquotes}
\usepackage{tikz} 
\usetikzlibrary{calc,arrows,decorations.markings} 
\usepackage{pict2e}
\usepackage{pgfplots}
\usepackage{float}
\usepackage{hyperref}
\usepackage{xr}
\usepackage{rotating}
\usepackage{amsmath,amssymb}
\usepackage{fp}
\usetikzlibrary{calc} 
\usepackage{booktabs, tabularx}
\usepackage{romannum}
\usepackage{bm}
\usepackage{float}
\usepackage{adjustbox}
\usepackage{booktabs, tabularx}
\usepackage{longtable}

\newcommand{\legLine}[5]{ 
			     \coordinate (center1) at (#1,#2); 
                \coordinate (b) at ($ (center1) - (.075,0) $);
                \coordinate (c) at ($ (b) - (.15,0) $); 
				\node at ($ (center1) + (#5,0.025) $) {{\color{#3}#4}}; 
                \draw[#3][line width=0.1mm] (b) -- (c); 
}

\newcommand{\legSqTxt}[3]{
		\begin{tikzpicture}
			     \coordinate (center1) at (#1,#2); 
                \coordinate (b) at ($ (center1) - (.075,0) $);
                \coordinate (c) at ($ (b) - (.1,0) $); 
                \coordinate (d) at ($ (center1) + (.075,0) $);
                \coordinate (e) at ($ (d) + (.1,0) $); 
				 \draw[#3,fill=#3] ($(b)-(0.0,0.075)$) rectangle ($ (d) + (0.0,0.075) $); 
		\end{tikzpicture}		
}

\newcommand{\circTxtFill}[3]{
		\begin{tikzpicture}
                \coordinate (center1) at (#1,#2); 
				 \draw[#3,fill=#3] (center1) circle (2pt);
		\end{tikzpicture}		
}

\newcommand{\legDiamondTxt}[3]{
		\begin{tikzpicture}
			     \coordinate (center1) at (#1,#2); 
                \coordinate (b) at ($ (center1) + (0.075,0.0) $); 
                \coordinate (c) at ($ (center1) + (0,0.075) $); 
                \coordinate (d) at ($ (center1) + (-0.075,0) $);
                \coordinate (e) at ($ (center1) + (0.0,-0.075) $);
				 \draw[#3,fill=#3] (b) -- (c) -- (d) -- (e) -- (b);
                \coordinate (c) at ($ (b) + (.1,0) $); 
                \coordinate (e) at ($ (d) - (.1,0) $); 
		\end{tikzpicture}		
}
                
\newcommand{\legTriangleTxt}[3]{
		\begin{tikzpicture}
			     \coordinate (center1) at (#1,#2); 
                \coordinate (b) at ($ (center1) - 0.17*(0.5,0.289) $); 
                \coordinate (c) at ($(center1) + 0.17*(0.5,-0.289) $); 
                \coordinate (d) at ($ (center1) + 0.17*(0,0.366) $);
				\draw[black,fill=#3, line width=0.1mm] 
    (b) -- (c) -- (d) -- cycle;                 
    			\coordinate (b) at ($ (center1) - (.075,0) $);
                \coordinate (c) at ($ (b) - (.1,0) $); 
                \coordinate (d) at ($ (center1) + (.075,0) $);
                \coordinate (e) at ($ (d) + (.1,0) $); 
		\end{tikzpicture}		
}		
		
\definecolor{light-gray}{gray}{0.85}

\newcommand{\leftTriang}[3]{
		\begin{tikzpicture}
			     \coordinate (center1) at (#1,#2); 
                \coordinate (b) at ($ (center1) + 0.17*(0.289,0.5) $); 
                \coordinate (c) at ($ (center1) + 0.17*(0.289,-0.5) $); 
                \coordinate (d) at ($ (center1) + 0.17*(-0.366,0.0) $);
				 \draw[fill=#3,#3] (b) -- (c) -- (d) -- (b);
                \coordinate (b) at ($ (center1) - (.075,0) $);
                \coordinate (c) at ($ (b) - (.1,0) $); 
                \coordinate (d) at ($ (center1) + (.075,0) $);
                \coordinate (e) at ($ (d) + (.1,0) $); 
		\end{tikzpicture}		
}

\newcommand{\rightTriang}[3]{
		\begin{tikzpicture}
			    \coordinate (center1) at (#1,#2); 
                \coordinate (b) at ($ (center1) + 0.17*(-0.289,0.5) $); 
                \coordinate (c) at ($ (center1) + 0.17*(-0.289,-0.5) $); 
                \coordinate (d) at ($ (center1) + 0.17*(0.366,0.0) $);
				 \draw[black,fill=#3] (b) -- (c) -- (d) -- (b);
                \coordinate (b) at ($ (center1) - (.075,0) $);
                \coordinate (c) at ($ (b) - (.1,0) $); 
                \coordinate (d) at ($ (center1) + (.075,0) $);
                \coordinate (e) at ($ (d) + (.1,0) $); 
		\end{tikzpicture}		
 }
\newcommand{\drawSlope}[6]{ 
				\coordinate (center1) at (#1,#2); 
				 \FPeval{\nx}{cos(#4*pi/180)}%
				 \FPeval{\ny}{sin(#4*pi/180)}%
				 \FPeval{\absNx}{abs(\nx)}%
				 \FPeval{\absNy}{abs(\ny)}%
				\coordinate (b) at ($ (center1) + #3*(-\ny,+\nx) $);
				\coordinate (c) at ($ (center1) - #3*(-\ny,+\nx) $); 
				 \FPeval{\xx}{(-\nx*\ny)}%
				\ifdim\xx pt < 0pt 
				\coordinate (d) at ($ (c) -2*#3*\ny*(1,0) $);
				\node at ($(d) +(0,#3*\nx)-0.2*(\nx/\absNx,0)$) {{\color{#5}#6}}; 
				\else
				\coordinate (d) at ($ (c) +2*#3*\nx*(0,1) $);
				\node at ($(d)-#3*\nx*(0,1)-0.2*\nx/\absNx*(1,0)$) {{\color{#5}#6}}; 
				\fi
				\draw[#5,line width=0.1mm] (d) -- (b); 
				\draw[#5,line width=0.1mm] (d) -- (c); 
				\draw[#5,line width=0.1mm] (b) -- (c); 
} 

\newcommand*{\Labelxy}[4]{\put(#1,#2) {\setlength{\fboxsep}{0pt}{\strut\textcolor{black}{\begin{turn}{#3}{#4}\end{turn}}}}}
\newcommand*{\LabelFig}[3]{\put(#1,#2) {\setlength{\fboxsep}{0pt}\colorbox{white}{\textcolor{black}{#3}}} }
\definecolor{darkolivegreen}{rgb}{0.33, 0.42, 0.18}
\definecolor{darkspringgreen}{rgb}{0.09, 0.45, 0.27}
\definecolor{darkslategray}{rgb}{0.18, 0.31, 0.31}
\definecolor{darkred}{rgb}{0.55, 0.0, 0.0}
\addeditor{KK}

\newcommand{\glfive}{Co$_5$Cr$_2$Fe$_{40}$Mn$_{27}$Ni$_{26}$~}
\newcommand{\glfour}{CoNiCrFeMn~}
\newcommand{\glthree}{CoCrFeMn~}
\newcommand{\gltwo}{CoNiCrFe~}
\newcommand{\glone}{CoNiFe~}
\newcommand{\glzero}{FeNi~}

\newcommand{\pmax}{$p_{\text{max}}$~}
\newcommand{\gmax}{\gamma_{\text{max}}}
\newcommand{\hmin}{$h_{\text{min}}$~}
\usepackage{xr}
\makeatletter
\newcommand*{\addFileDependency}[1]{
  \typeout{(#1)}
  \@addtofilelist{#1}
  \IfFileExists{#1}{}{\typeout{No file #1.}}
}
\makeatother
\newcommand*{\myexternaldocument}[1]{
    \externaldocument{#1}
    \addFileDependency{#1.tex}
    \addFileDependency{#1.aux}
}
\myexternaldocument{./sm}

\begin{document}

\title{\change{Percolation of elastic modulus heterogeneities in  multi-component metallic glasses near plastic yielding}{Yielding in multi-component metallic glasses: Universal signatures of elastic modulus heterogeneities}}

\author{Kamran Karimi$^1$}
\email{Corresponding author: kamran.karimi@ncbj.gov.pl}
\author{Mikko J. Alava$^{1,2}$}
\author{Stefanos Papanikolaou$^1$}
\affiliation{%
 $^1$ NOMATEN Centre of Excellence, National Center for Nuclear Research, ul. A. Sołtana 7, 05-400 Swierk/Otwock, Poland\\
 $^{2}$ Aalto University, Department of Applied Physics, PO Box 11000, 00076 Aalto, Espoo, Finland
}%

\begin{abstract}
Sheared multi-component bulk metallic glasses are characterized by both chemical and structural disorder that define their properties. We investigate the behavior of the local, microstructural elastic modulus across the plastic yielding transition in six Ni-based multi-component glasses, that are characterized by compositional features commonly associated with solid solution formability. We find that elastic modulus fluctuations display consistent percolation characteristics pointing towards universal behavior across chemical compositions and overall yielding sharpness characteristics. Elastic heterogeneity grows upon shearing via the percolation of elastically soft clusters within an otherwise rigid amorphous matrix, confirming prior investigations in granular media and colloidal glasses. We find \change{varying degrees}{clear signatures} of percolation transition with spanning clusters that are universally characterized by scale-free characteristics and critical scaling exponents. 
The spatial correlation length and mean cluster size tend to diverge prior to yielding, with associated critical exponents that exhibit fairly weak dependence on compositional variations as well as macroscopic stress-strain curve details.
\end{abstract}

\maketitle
\remove{Shear banding instability in a driven metallic glass generically occurs via localization of intense (irrecoverable) deformation that the glassy material undergoes beyond its elastic limit without any cracking or crumbling. 
The microscopic basis of this \change{viewpoint}{failure transition} is \add{believed to be} the emergence of a \emph{collective} dynamics mediated by Shear Transformation Zones (STZs) \cite{argon1979plastic,falk1998dynamics, falk2011deformation}.  
Owing to their (mechanical) softness, STZs are commonly considered as fertile sites to undergo localized plastic rearrangements within the amorphous matrix but with long-range (compared to their own size) elastic-type consequences \cite{eshelby1959elastic,picard2004elastic,karimi2019plastic}.
In a mechanically driven glass, these soft spots are initially activated by external deformations, but further instability may be triggered and propagate due to non-local interactions.
Near yielding transition, the so-called \emph{avalanche dynamics} may emerge in which the activation process takes place by sequentially forming STZ clusters of all scales (but bounded to the physical scale limit) \cite{karimi2017inertia}.  
In this regard, plastic yielding may be viewed as a non-equilibrium phase transition with unique characteristics, such as diverging length and/or timescales and power-law distributions of avalanche sizes, associated with it \cite{zapperi1997plasticity,ParisiPNAS2017}.}

In realistic studies of bulk metallic glasses, the associated yielding properties and shear banding behavior \cite{dai2008basic} have important industrial and technological implications in terms of the \emph{ductility} of the glass.
The ongoing eagerness to promote the latter (in combination with hardening properties) have already shifted the focus to the development of \emph{multi-component} glasses (also known as amorphous alloys and/or high-entropy metallic glasses) with the highly tunable microstructural/compositional complexity \cite{schuh2007mechanical,ding2013high}. 
The atomistic origin of this new design paradigm is based on the common observation that glasses with low atomic disparity limit (where atomic radii of constituent elements are considered to be very similar) have a high capacity to form localized deformation patterns whereas those at the opposite limit have a tendency to delocalize strain \cite{karimi2021shear} and, therefore, deform in a more ductile way.
In this context, the shear band structure and design-level ductility appear to be highly dependent on inherent heterogeneity as an essential elemental/compositional feature.
However, microstructural origins and elemental dependence of such inhomogeneities and, more importantly, associated spatial-dynamical evolution upon shearing have yet to be explored.

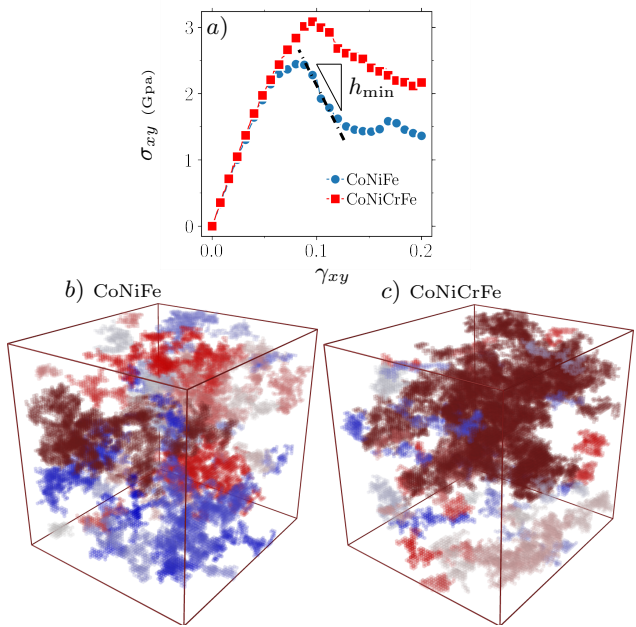
\begin{figure}[t]
    \centering
    \begin{overpic}[width=0.23\textwidth]{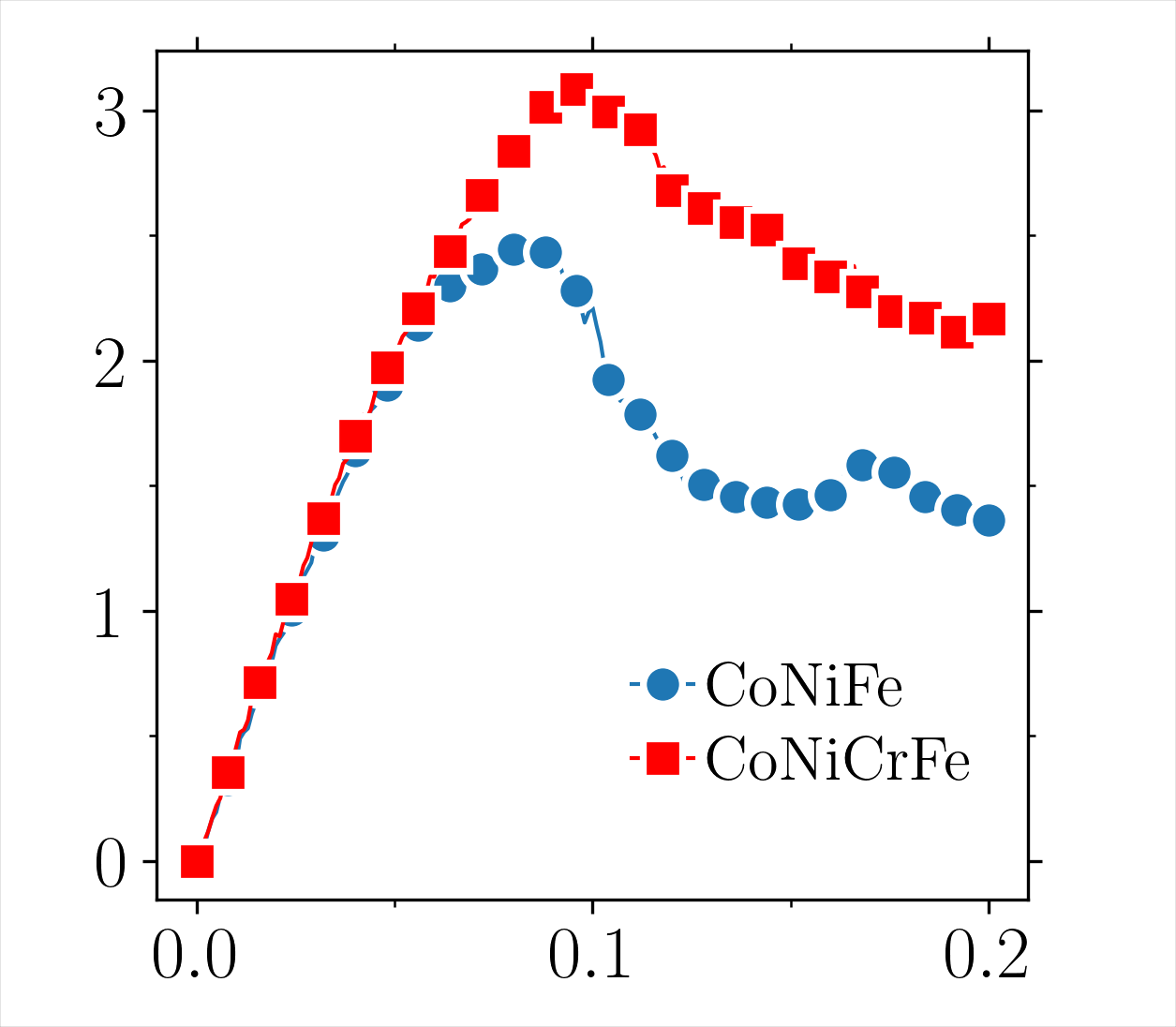}
        \LabelFig{15}{76}{$a)$}
        \Labelxy{50}{-3}{0}{$\gamma_{xy}$}
        \Labelxy{-6}{35}{90}{{$\sigma_{xy}$ \tiny(Gpa)}}
        %
       \begin{tikzpicture}
            \coordinate (a) at (0,0); 
            \node[white] at (a) {\tiny.};               %
                 \drawSlope{2.1}{2.3}{0.35}{208}{black}{\hspace{12pt}$h_\text{min}$}
			\coordinate (center2) at (2,2.2); 
            \coordinate (b) at ($ (center2) - .25*(1.2,-2.4) $); 
            \coordinate (c) at ($ (center2) + .25*(1.2,-2.4) $);
            \draw[black,dash dot,thick][line width=0.4mm] (c) -- (b); 
 		    \end{tikzpicture}
     \end{overpic}
    \vspace{12pt}

    \begin{overpic}[width=0.23\textwidth]{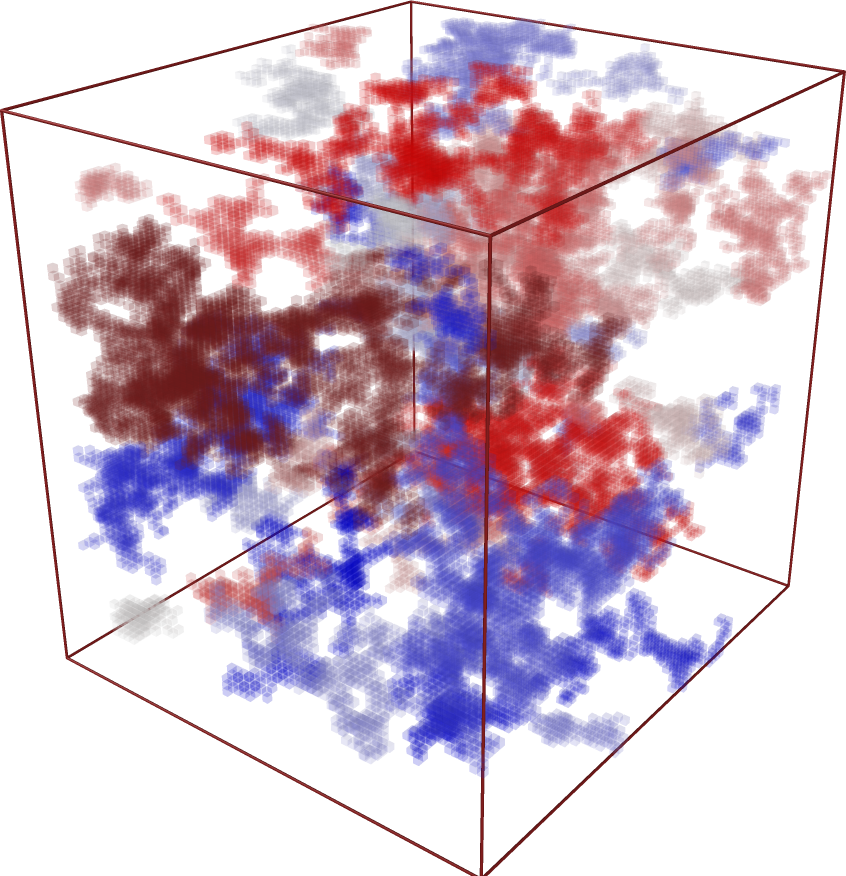}
        \LabelFig{18}{102}{$b)$ \scriptsize \glone}
    \end{overpic}
    \begin{overpic}[width=0.23\textwidth]{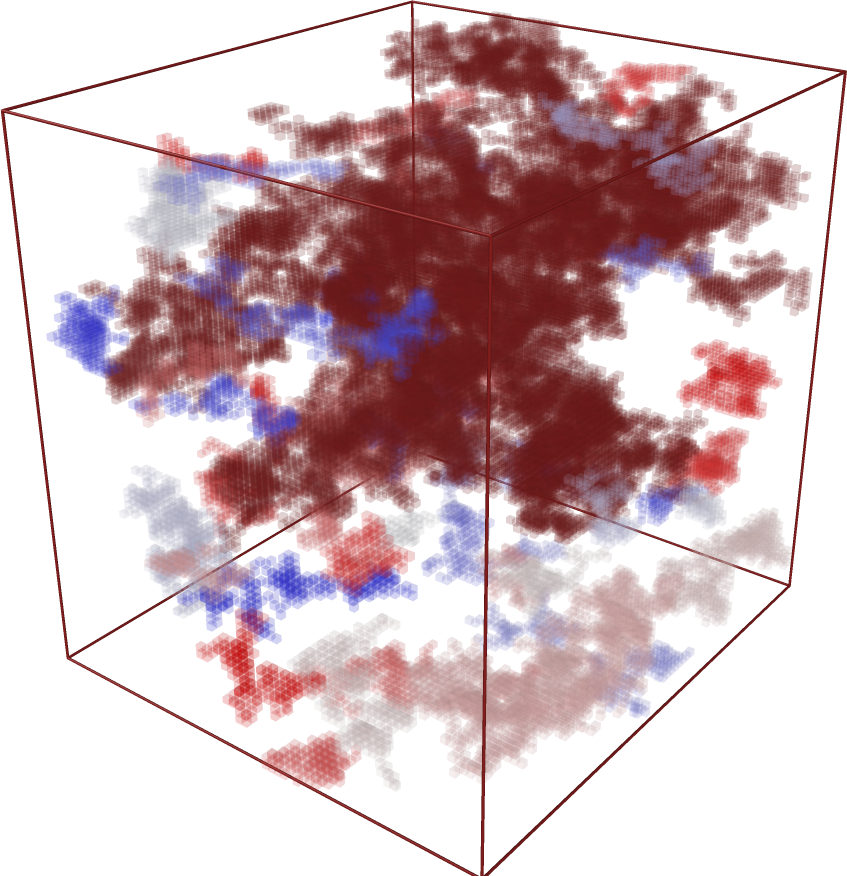}
        \LabelFig{18}{102}{$c)$ \scriptsize \gltwo}
    \end{overpic}
    %
%
    \caption{Three dimensional softness maps of the \textbf{b}) \glone \textbf{c}) \gltwo compositions at applied shear strain $\gamma_{xy}\simeq 0.1$ and the corresponding stress-strain curves $\sigma_{xy}$ versus $\gamma_{xy}$ in \textbf{a}). The colors in \textbf{b}) and \textbf{c}) indicate different clusters including elastically unstable regions with $\mu\le 0$. The slope in \textbf{a}) indicates the softening modulus \hmin. Upon shearing, negative-$\mu$ clusters form percolating networks (in brown) on approach to failure that tend to correlate with the (post-failure) macroscopic properties.}
    \label{fig:loadCurve}
\end{figure}

Despite certain commonalities of failure properties across disordered solids \cite{denisov2017universal}, the \emph{sharpness} of yielding transition, \change{i.e. the extent}{as a common signature} of strain localization (and ductility), may show strong variations by altering thermal treatments and chemical compositions in bulk metallic glasses, ranging from uniformly distributed patterns to system-spanning crack-like features \cite{schuh2007mechanical,karimi2021shear,cheng2011atomic,cheng2009correlation}.
Owing to the presence of quenched chemical/structural disorder, glassy metals may accommodate a distributed plastic flow  with a significant contribution to ductility \cite{wang2018spatial,wu2009transition,chen2006extraordinary}.
Certain (aged) glasses that lack this heterogeneity element \cite{cheng2008local,shi2005strain,albano2005shear} (or associated lengths don't exceed interatomic scales) tend to localize strain within a single dominant band before shear instablility results in a catastrophic brittle-type fracture.
In this context, tailoring \emph{elastic} heterogeneity has recently emerged as a novel design framework to build more ductile metallic glasses \cite{wang2018spatial,wang2018multi}.
Under special thermal treatments and variations of chemical elements, quenched metallic glasses may nucleate elastically soft clusters that become structured \remove{over mesoscopic lengths} \add{and \emph{span} the entire system upon failure}, leading to the enhanced plastic flow and, hence, ductility.
\add{This is illustrated in Fig.~\ref{fig:loadCurve} where yielding transition, fine-tuned by the chemical compositions, is accompanied by the percolating networks of \emph{softness} (shown in brown) in a driven \glone and \gltwo bulk metallic glass.}
\add{The relevance of such a percolation transition of softness within the context of amorphous plasticity has been already established in (soft) model glasses \cite{nagamanasa2014direct,shrivastav2016yielding,ghosh2017direct} as well as colloidal systems \cite{ghosh2017direct,schall2007structural} under shear.
These studies have mainly focused on certain (implicit) notions of softness based on the degree of deformation non-affinity or shear-induced mobility that exhibit percolating features near the criticality.
Using a model metallic glass in \cite{mayr2009relaxation}, the glass transition was accompanied by the percolation threshold of mechanically unstable regions on approach to the transition temperature $T_g$.
However, realistic studies of bulk metallic glasses driven out of equilibrium, with substantial industrial and technological implications, and associated composition-dependence has been more challenging.
}

Here, using atomistic simulations, we investigate six multi-component bulk metallic glasses including Co, Ni, Fe, Cr, and Mn at compositions that have been the focus of recent experimental investigations.
{The chosen compositions are commonly believed to optimize solid solution formability, characterized by low misfit coefficients $\delta_a$, and not thought to promote glass forming ability typically observed for $\delta_a>6\%$ \cite{zhang2008solid}.
Nonetheless, there is a way to generate a glassy environment in the opposite limit of $\delta_a \rightarrow 0$ but maintaining a larger number of elements, as in medium/high entropy metallic glasses (see \cite{ding2013high} and references therein).}

By probing local elasticity, in particular shear modulus $\mu$ and its evolution toward failure, we present direct evidence that plastic yielding in driven metallic glasses takes place \emph{universally} through a percolation transition of softness, characterized by regions of negative $\mu \le 0$ (as shown in Fig.~\ref{fig:loadCurve}).
The sharpness of yielding transition, fine-tuned by chemical compositions and inferred from the rate of spontaneous drop \hmin in the stress response, exhibits \emph{consistent} correlation features with softness and its percolation properties. 
Our cluster analysis associated with local elasticity maps features critical scaling properties, i.e. scale-free statistics, divergence of correlation lengths, and critical exponents that are reminiscent of a non-equilibrium phase transition.

From a broader perspective, our methodology offers a robust indicator of nonaffinity, e.g. the disorder-induced breakdown of homogeneous response, by probing elastic heterogeneity in amorphous materials.
Conceptually, our approach is similar to investigations of low energy quasi-localized modes that are spatially distinguishable from long wave-length phonon modes in elastic crystalline solids (see \cite{richard2020predicting} and references therein).
Predictive plasticity models \emph{solely} based on the notion of shear transformation zones (STZs) might fail to capture such correlations because the latter are generally believed to be (micro)structural defects but not necessarily indicative of mechanical instability.
Another complication arises from a lack of robust topological signatures that can identify STZs from their parent liquid-like microstructure within a glassy matrix.
There have been related efforts based upon the machine learning-rooted concept of ``structural" softness but with very limited predictive capabilities in terms of glass failure and deformation \cite{cubuk2015identifying,fan2020machine,boattini2020autonomously}.
More focused studies of metallic glasses made an attempt to associate soft spots to the presence/absence of geometrically unfavored motifs and/or short/medium range ordering \cite{ding2014soft,ding2014full,ding2016universal} but theses local motifs on their own often fail to describe the collective nature of plasticity and shear banding in a broader context of amorphous plasticity.
Our study aims to augment such efforts, directed mainly towards the notion of structural heterogeneity, by adding the elastic heterogeneity picture to the scene which is more conforming with the long-established notion of structure-property correlations in (metallic) glasses.\\

\begin{figure}[b]
    \centering
    \begin{overpic}[width=0.23\textwidth]{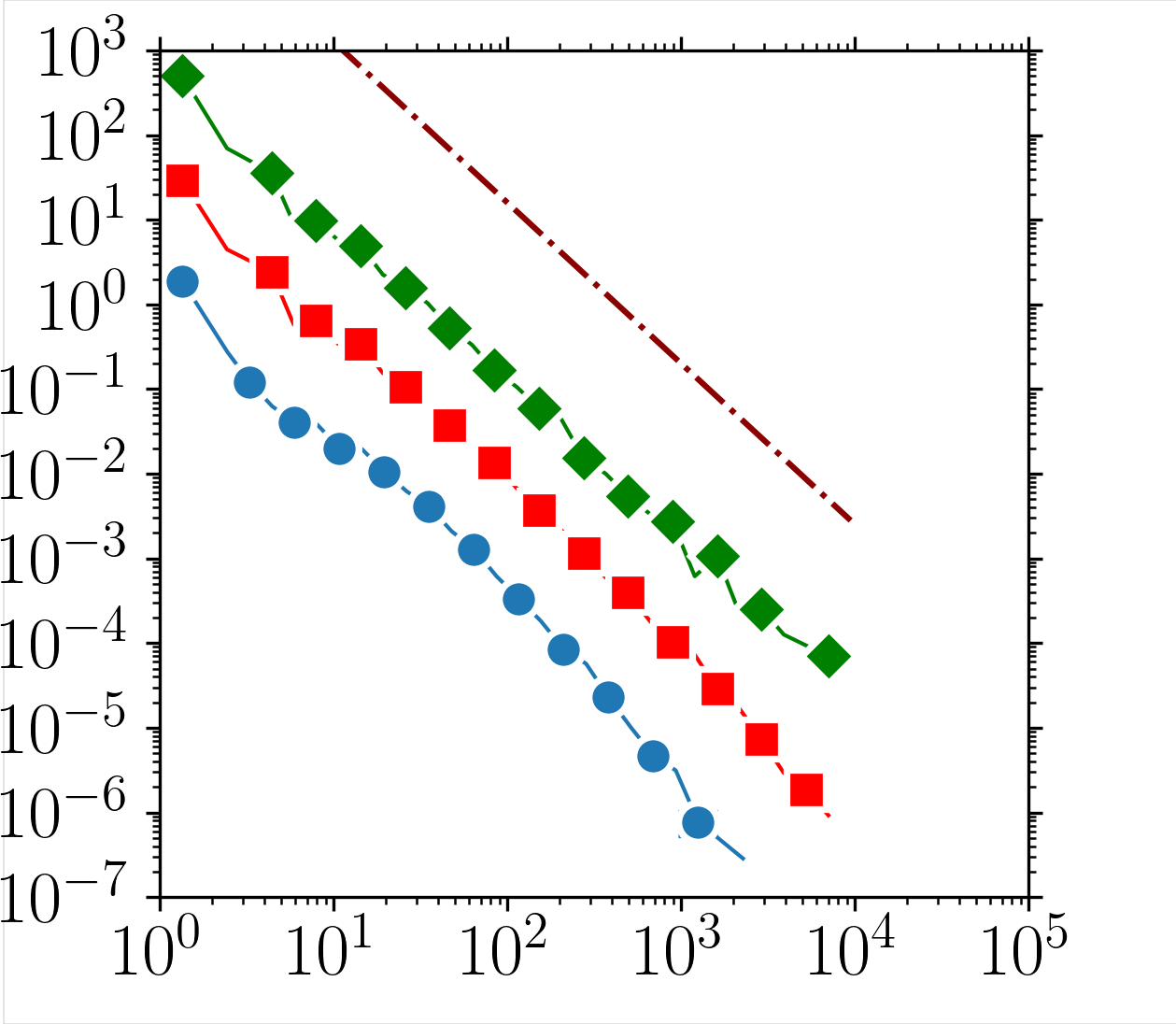}
        \LabelFig{16}{87}{$a)$ \scriptsize \glone}
        \Labelxy{50}{-2}{0}{$s$}
        \Labelxy{-8}{45}{90}{{$n_s$}}
        \begin{tikzpicture}
          \coordinate (a) at (0,0);
            \node[white] at (a) {\tiny.};               %
            \drawSlope{1.85}{2.7}{0.35}{230}{darkred}{$\tau$}
            \legLine{0.7}{0.5}{darkred}{$\scriptstyle n_s\propto s^{-\tau}$}{0.5}
		\end{tikzpicture}
	\end{overpic}
    \begin{overpic}[width=0.23\textwidth]{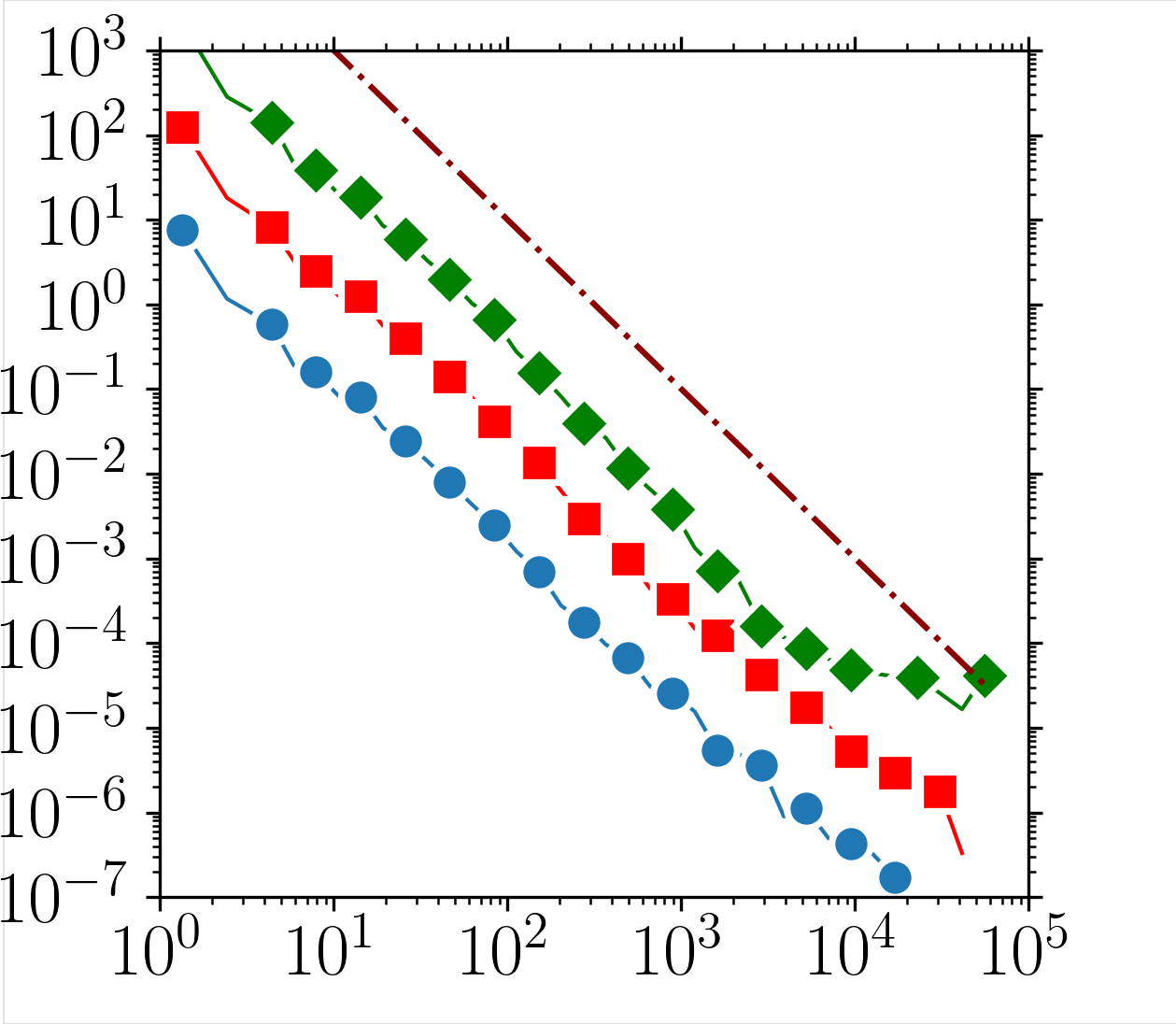}
        \LabelFig{16}{87}{$c)$  \scriptsize \gltwo}
        \Labelxy{50}{-2}{0}{$s$}
        \Labelxy{-8}{45}{90}{{$n_s$}}
        \begin{tikzpicture}
          \coordinate (a) at (0,0);
            \node[white] at (a) {\tiny.};               %
            \drawSlope{1.8}{2.7}{0.35}{229}{darkred}{$\tau$}
            \legLine{0.7}{0.5}{darkred}{$\scriptstyle n_s\propto s^{-\tau}$}{0.5}
		\end{tikzpicture}
	\end{overpic}
    \begin{overpic}[width=0.23\textwidth]{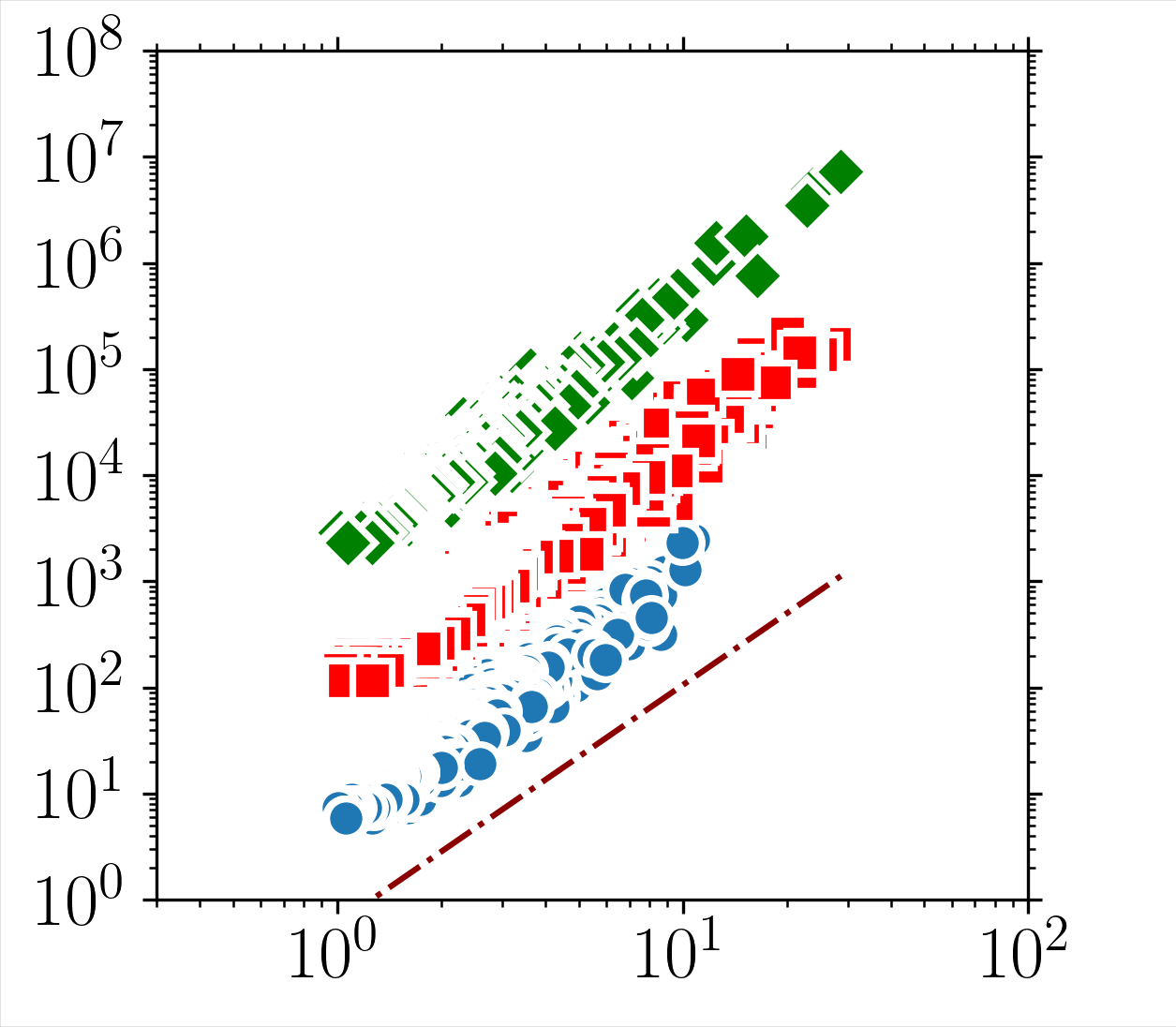}
        \LabelFig{16}{16}{$b)$}
        \Labelxy{50}{-6}{0}{$r_s$}
        \Labelxy{-8}{45}{90}{{$s$}}
        %
        \begin{tikzpicture}
          \coordinate (a) at (0,0);
            \node[white] at (a) {\tiny.};               %
            \drawSlope{2.15}{.9}{0.35}{126}{darkred}{$d_f$}
            \legLine{0.8}{3}{darkred}{$\scriptstyle s\propto r_s^{d_f}$}{0.4}
		\end{tikzpicture}
	\end{overpic}
    \begin{overpic}[width=0.23\textwidth]{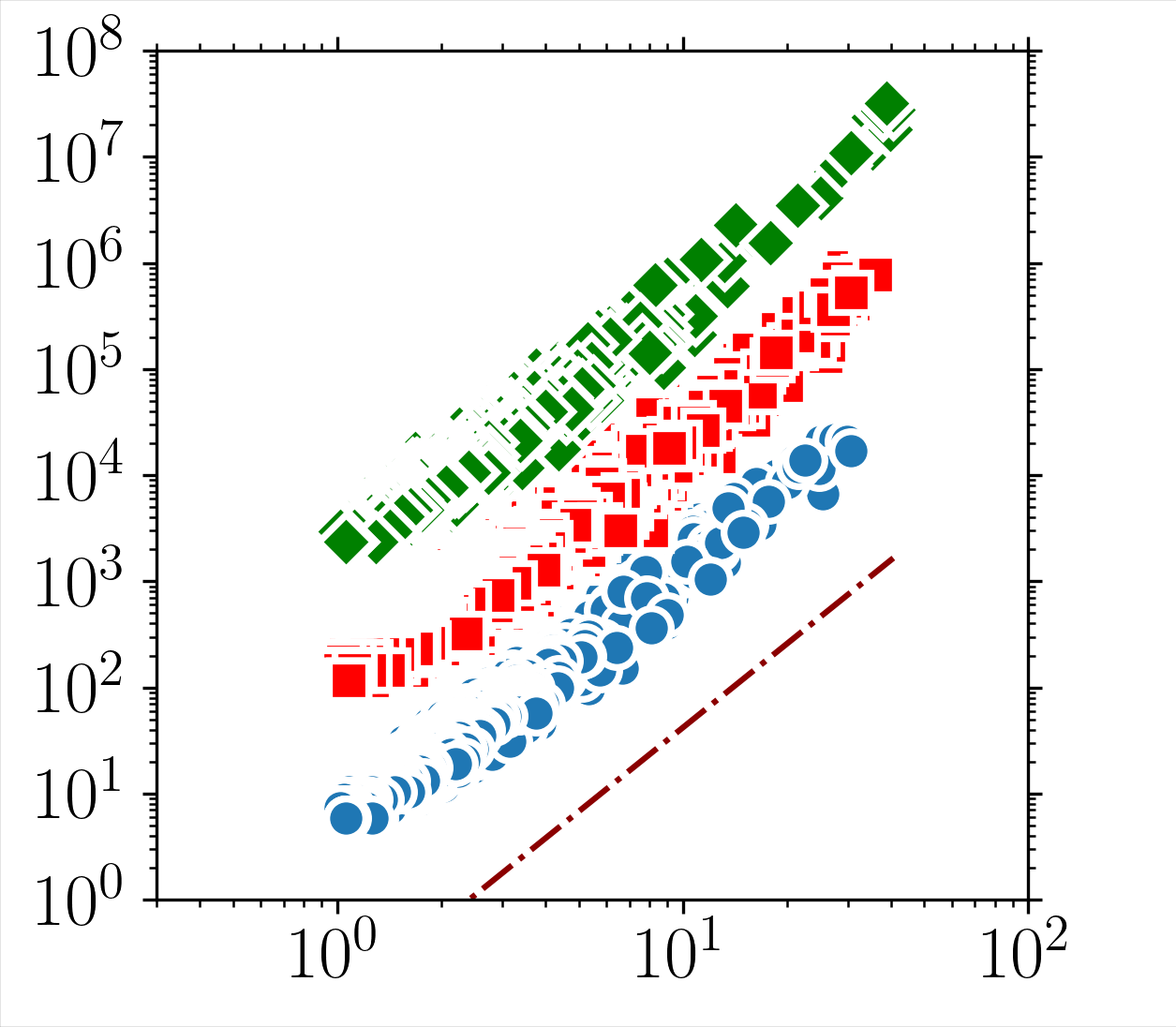}
        \LabelFig{16}{16}{$d)$}
        \Labelxy{50}{-6}{0}{$r_s$}
        \Labelxy{-8}{45}{90}{{$s$}}
        %
        \begin{tikzpicture}
          \coordinate (a) at (0,0);
            \node[white] at (a) {\tiny.};               %
            \drawSlope{2.35}{.9}{0.35}{128}{darkred}{$d_f$}
            \legLine{0.8}{3}{darkred}{$\scriptstyle s\propto r_s^{d_f}$}{0.4}
		\end{tikzpicture}
	\end{overpic}
    \caption{Cluster size statistics corresponding to the \glone and \gltwo glasses. \textbf{a},\textbf{c}) Cluster size distribution $n_s$ \textbf{b},\textbf{d}) scatter plot of cluster size $s$ and associated radius of gyration $r_s$ at different srains $\gamma_{xy}=0.03$ (\protect\circTxtFill{0}{0}{blue}), $0.07$ (\protect\legSqTxt{0}{0}{red}), $0.1$ (\protect\legDiamondTxt{0}{0}{darkspringgreen}). The dashdotted lines denote power laws $n_s\propto s^{-\tau}$ with \textbf{a}) $\tau=2.0$ \textbf{c}) $\tau=2.1$ and $s\propto r_s^{d_f}$ with \textbf{b}) $d_f=2.3$ \textbf{d}) $d_f=2.6$.} 
    \label{fig:clusterSize}
\end{figure}
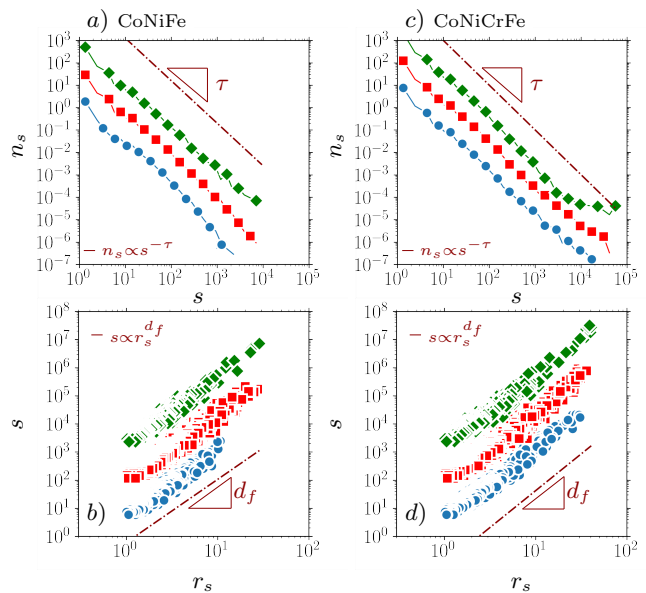

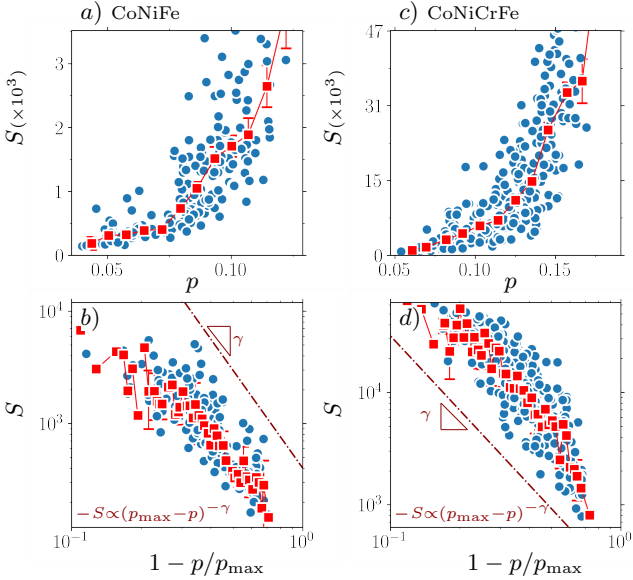
\begin{figure}[t]
    \begin{overpic}[width=0.23\textwidth]{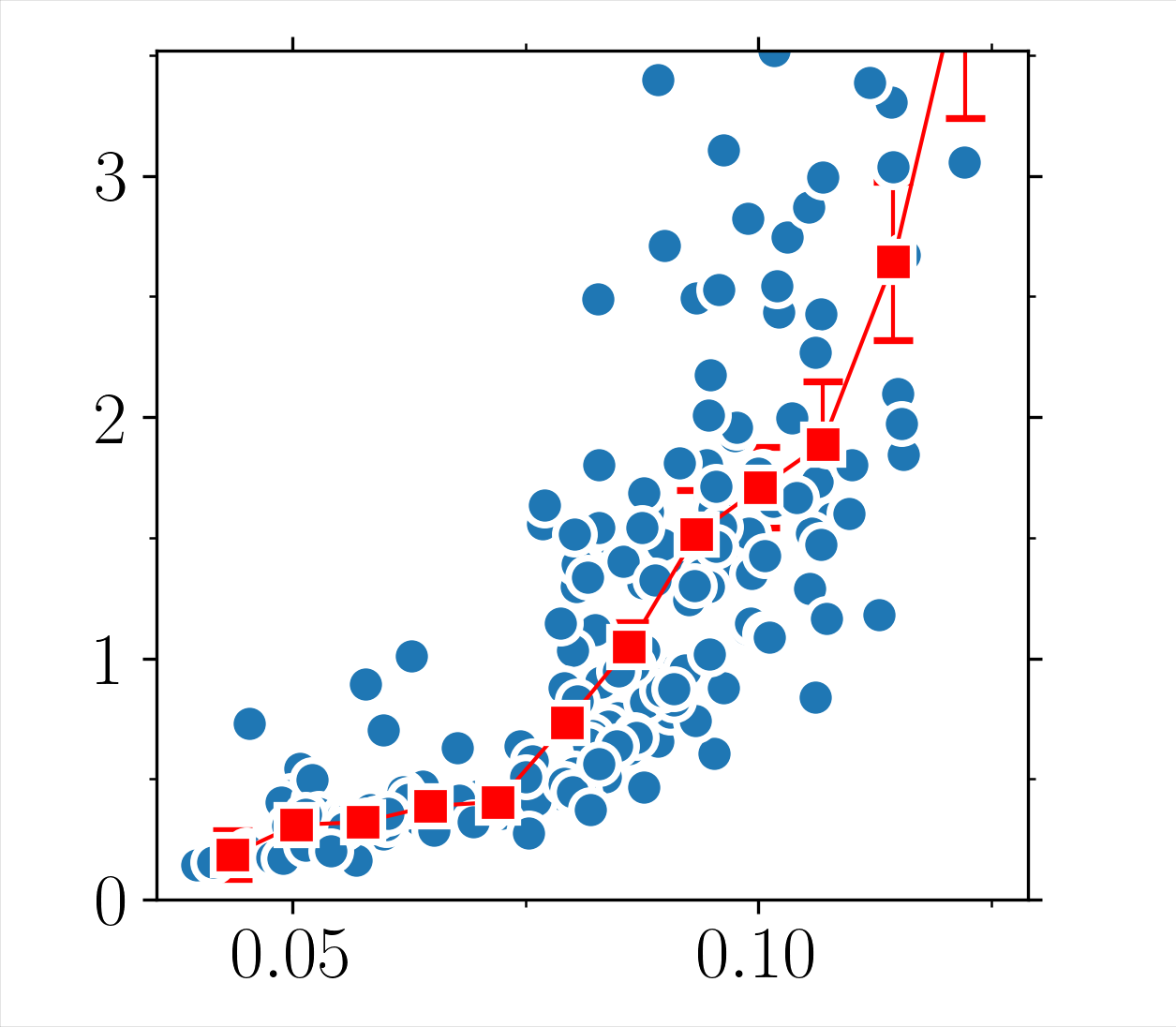}
        \LabelFig{16}{87}{$a)$ \scriptsize \glone}
        \Labelxy{50}{0}{0}{$p$}
        \Labelxy{-8}{46}{90}{{$S\scriptstyle(\times 10^{3})$}}
    \end{overpic}
    \begin{overpic}[width=0.23\textwidth]{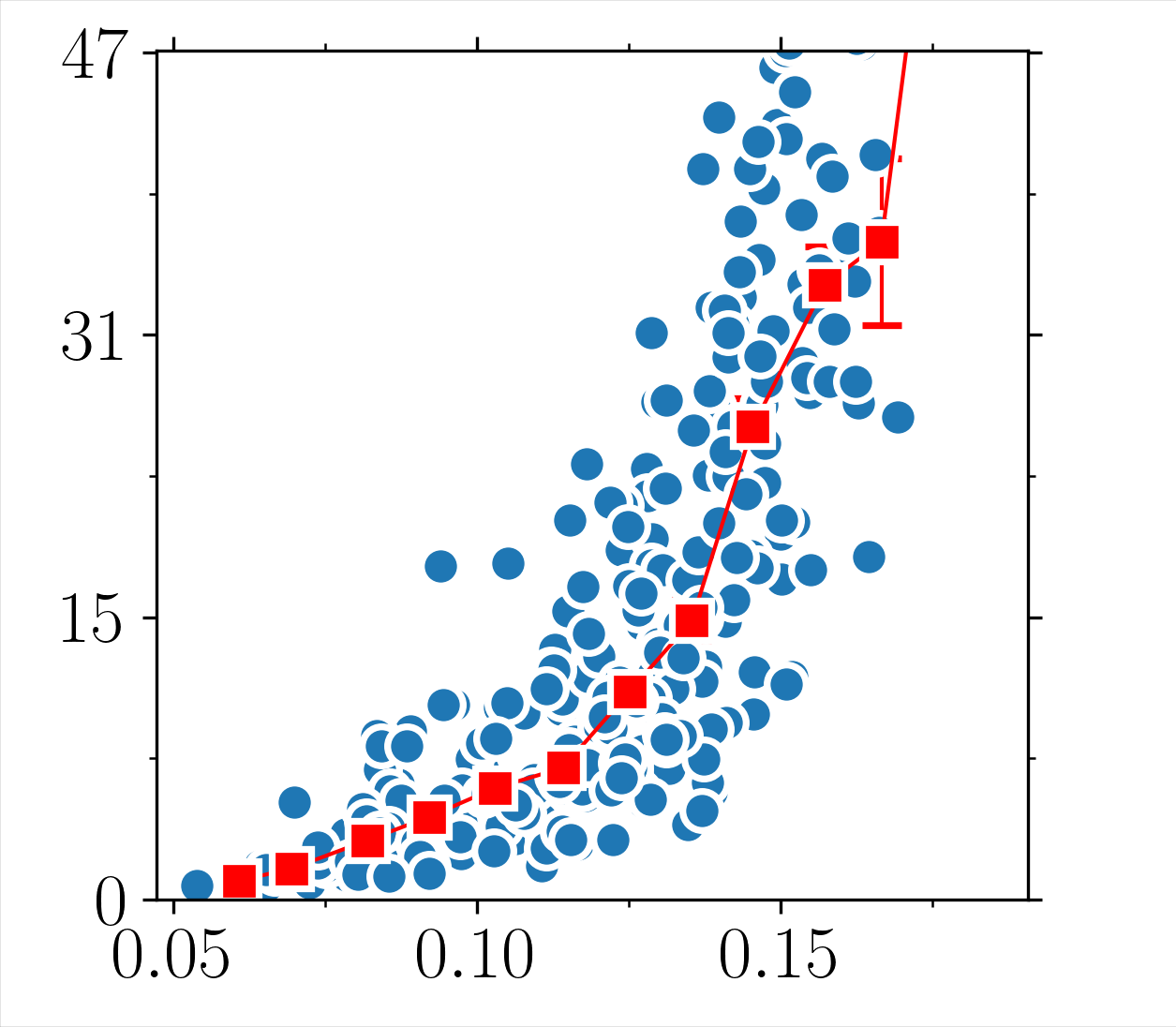}
        \LabelFig{16}{87}{$c)$ \scriptsize \gltwo}
        \Labelxy{50}{0}{0}{$p$}
        \Labelxy{-8}{46}{90}{{$S\scriptstyle(\times 10^{3})$}}
    \end{overpic}

    \begin{overpic}[width=0.23\textwidth]{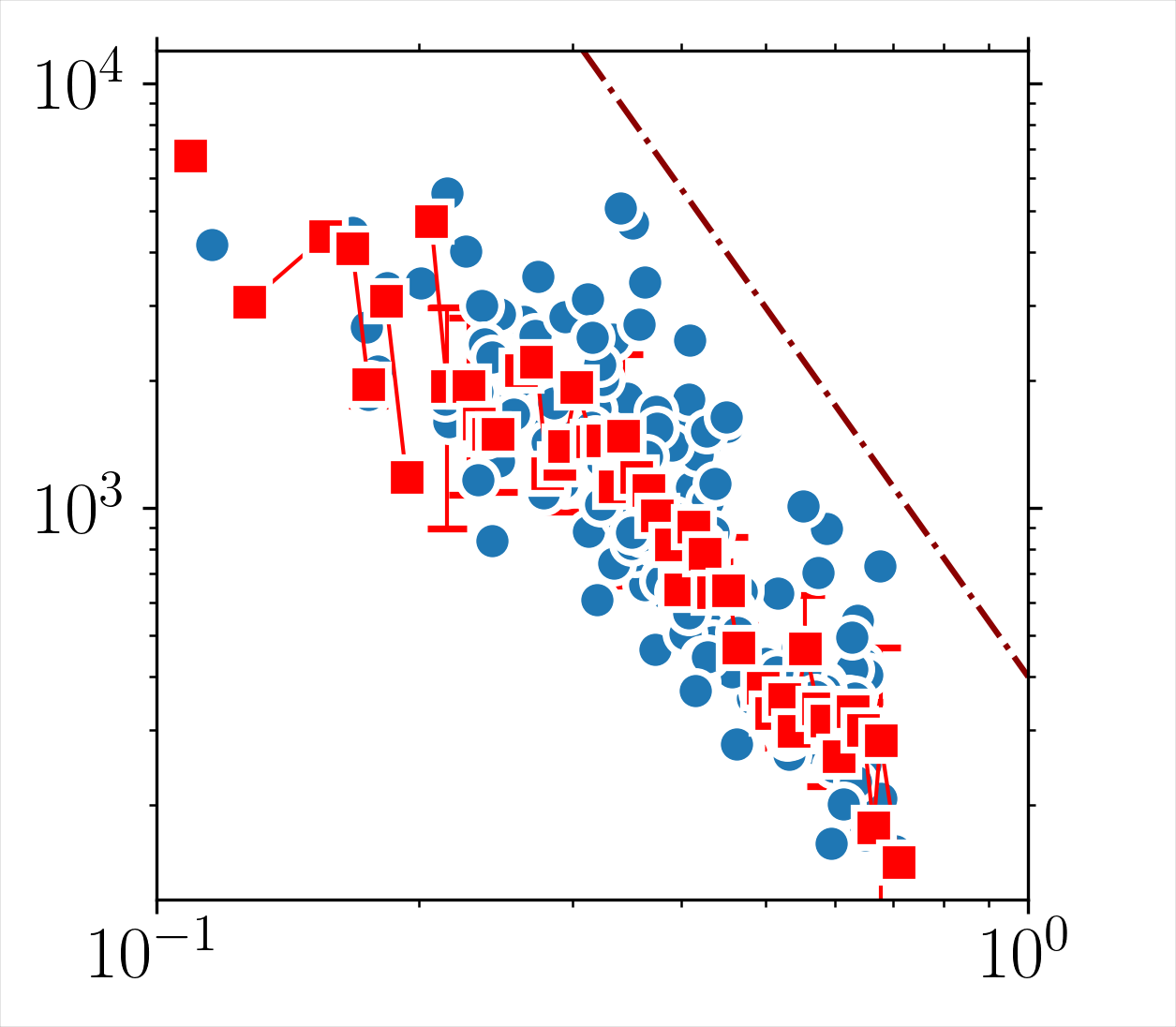}
        \LabelFig{16}{76}{$b)$}
        %
        \Labelxy{-8}{46}{90}{{$S$}}
        \Labelxy{38}{-3}{0}{$1-p/p_\text{max}$}
        \begin{tikzpicture}
          \coordinate (a) at (0,0);
            \node[white] at (a) {\tiny.};               %
            \drawSlope{2.35}{2.8}{0.25}{218}{darkred}{$\hspace{-5pt}\scriptstyle\gamma$}
            \legLine{0.7}{0.5}{darkred}{$\scriptstyle S\propto (p_{\text{max}}-p)^{-\gamma}$}{0.9}
		\end{tikzpicture}
    \end{overpic}
    \begin{overpic}[width=0.23\textwidth]{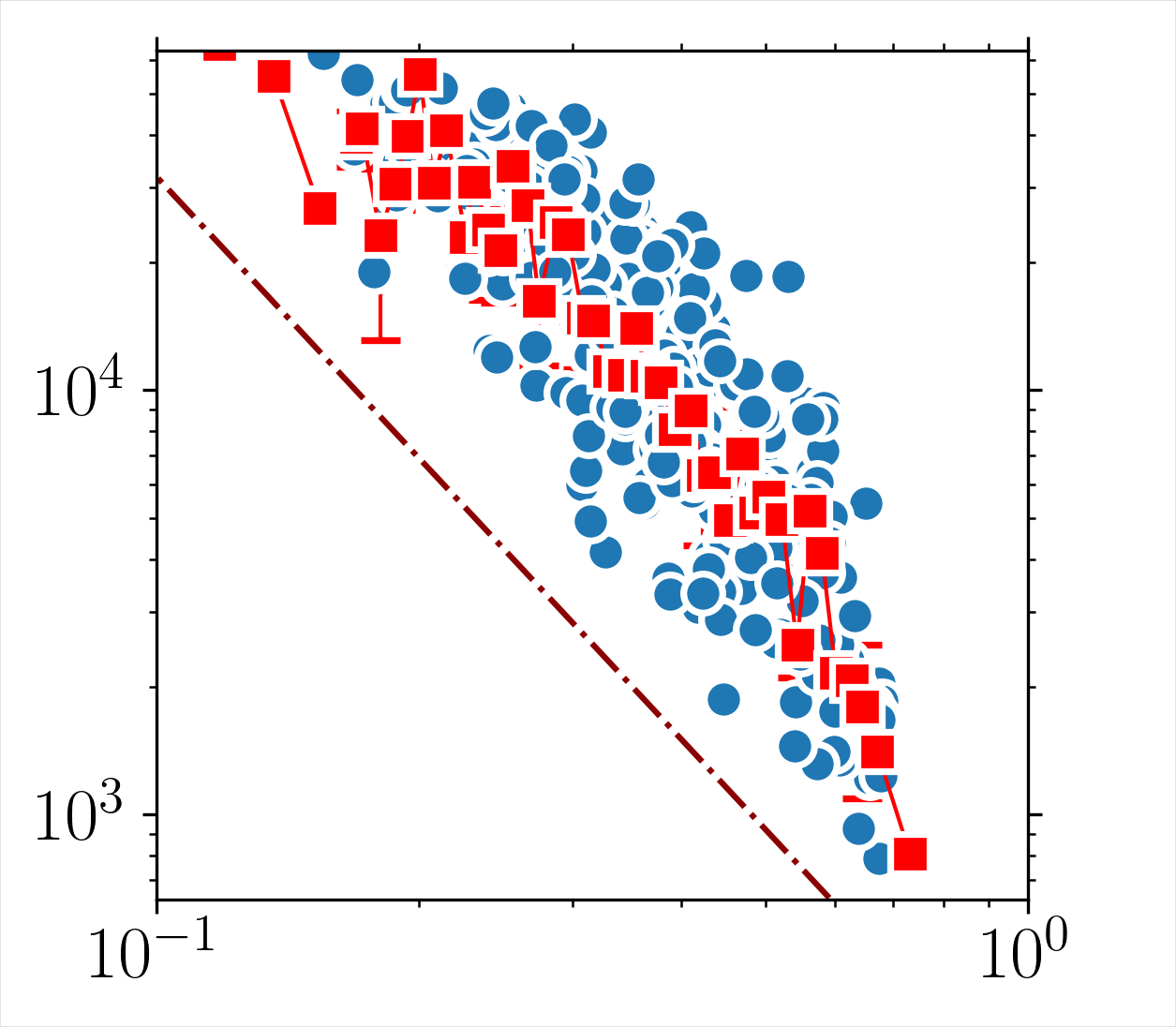}
        \LabelFig{16}{76}{$d)$}
        %
        \Labelxy{-8}{46}{90}{{$S$}}
        \Labelxy{38}{-3}{0}{$1-p/p_\text{max}$}
        \begin{tikzpicture}
          \coordinate (a) at (0,0);
            \node[white] at (a) {\tiny.};               %
            \drawSlope{1.25}{1.8}{0.25}{45}{darkred}{$\scriptstyle\gamma$}
            \legLine{0.7}{0.5}{darkred}{$\scriptstyle S\propto (p_{\text{max}}-p)^{-\gamma}$}{0.9}
		\end{tikzpicture}
    \end{overpic}
    \caption{Mean cluster size $S$ plotted against $p$ associated with the \textbf{a}) \glone and \textbf{c}) \gltwo glasses. The graphs in \textbf{b}) and \textbf{d}) are the same as \textbf{a}) and \textbf{c}) but plot $S$ as a function of $1-p/p_\text{max}$ with \textbf{a}) $p_{\text{max}}= 0.14$ and \textbf{c}) $p_{\text{max}}= 0.20$. The (red) curves indicate binning averaged data. The dashdotted lines denote power laws $S\propto (p_{\text{max}}-p)^{-\gamma}$ with \textbf{b}) $\gamma=2.89$ and \textbf{d}) $\gamma=2.20$.}
    \label{fig:smean}
\end{figure}

\noindent\emph{Simulations \& Protocols---}
Details of molecular dynamics simulations are given in \cite{karimi2021shear} including relevant units, initial configurations, interatomic forces, and deformation parameters of \glzero, \glone, \gltwo, \glthree, \glfour, and \glfive model metallic glasses.
In order to compute the elasticity tensor locally for simulated glasses, simple shear tests were performed on the $xy$ plane at a fixed strain rate $\dot{\gamma}_{xy}=10^{-5}~\text{ps}^{-1}$ and temperature $T=300$ K up to a prescribed (pre-)strain $\gamma_{xy} \le 0.2$.  
In line with \cite{mizuno2013measuring,tsamados2009local}, we subsequently perturbed the simulation cell through six deformation modes in Cartesian directions $xyz$ and evaluated resulting differences in atom-wise stresses to construct the local elasticity tensor for each atom (see the appendix).
Relevant atom-based quantities were interpolated on a fine cubic grid to be used as input for our three dimensional cluster processing.
\note{move the next sentence up.}
This methodology allows us to probe the spatial-dynamical evolution of the local shear modulus $\mu=c_{xyxy}$ and associated percolation features near failure transition in sheared glasses. 
We further measure the softening modulus \hmin defined as the maximum \emph{rate} of the \emph{macroscopic} stress drop for all the different compositions.
The stress drop, typically defined as the difference between the overshoot stress and the subsequent flow stress, is associated with the initiation of a catastrophic shear band and has been used as an appropriate order parameter in model glass studies \cite{ozawa2018random,chen2011theory} showing meaningful variations with glass compositions and processing parameters \cite{cheng2008local}.
We note that in metallic glass simulations and/or experiments a robust measurement of the macroscopic drop is not always feasible due to the lack of a well-defined steady flow regime that is expected to follow the stress overshoot.
In a recent work \cite{karimi2021shear}, we established \hmin as a more robust experimentally-relevant indicator of shear banding and associated structural features. Here, we show that the variations in \hmin tend to correlate with the softness properties as inferred from local elasticity. \\

\noindent\emph{Results}--- Figure~\ref{fig:loadCurve} displays results of the shear tests performed on the quenched \glone and \gltwo glasses.
The cluster maps based on unstable regions with $\mu\le 0$ are visualized in Fig.~\ref{fig:loadCurve}(b) and (c).
We denote the fraction of unstable sites by (probability) $p$ which appears to evolve in a non-monotonic fashion, Fig.~\ref{fig:statistics_gamma}(b) and (c), with a peak value $p_\text{max}$ that almost coincides with that of the (bulk) shear stress $\sigma_{xy}$ (at $\gamma_{xy}\simeq 0.1$) as in Fig.~\ref{fig:loadCurve}(a).
Apart from variations in $p_\text{max}$, Fig.~\ref{fig:statistics_gamma}(a-f) shows similar trends for the evolution of $p$ with $\gamma_{xy}$ corresponding to the other compositions.
We notice $p>0$ even in unstrained quenched glasses (at $\gamma_{xy}=0$) owing to the distributed (but disconnected) networks of soft (Eshelby-like \cite{eshelby1959elastic}) inclusions within the glassy matrix.
\remove{Therefore, $\langle\mu\rangle > 0$ at small applied strains, indicating a \emph{bulk} stability (despite a non-zero $p$) below the glass transition temperature $T_g$ \cite{mayr2009relaxation}.}
\note{The observed reduction of $p$ past the yield point almost coincides with an upturn observed in $\langle \mu \rangle$ (see appendix) within the softening regime. might be indicative of finite size effects?}

\begin{figure}[b]
    \begin{overpic}[width=0.23\textwidth]{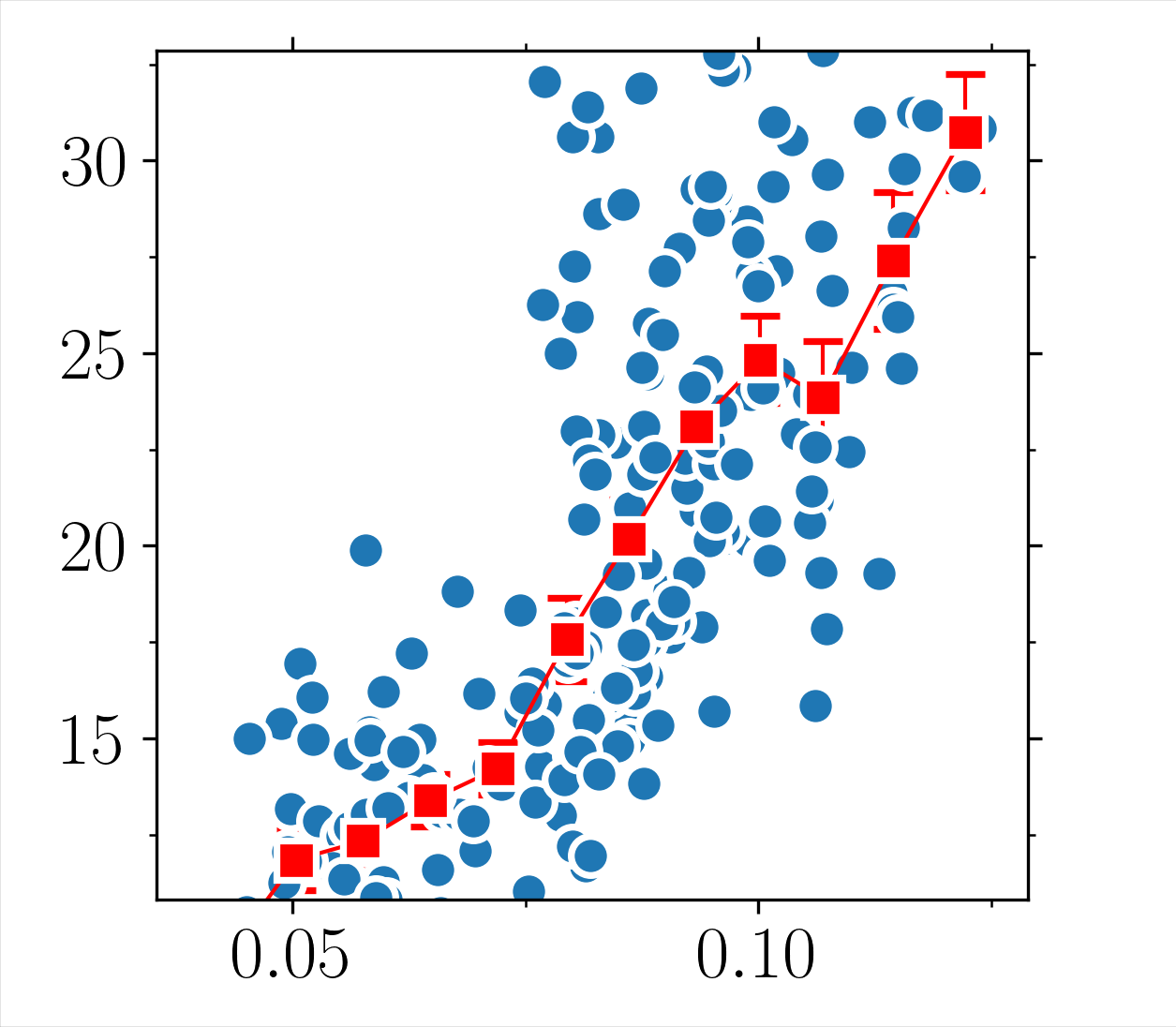}
        \LabelFig{16}{87}{$a)$ \scriptsize \glone}
        \Labelxy{50}{0}{0}{$p$}
        \Labelxy{-8}{46}{90}{{$\xi$\tiny (\r{A})}}
    \end{overpic}
    \begin{overpic}[width=0.23\textwidth]{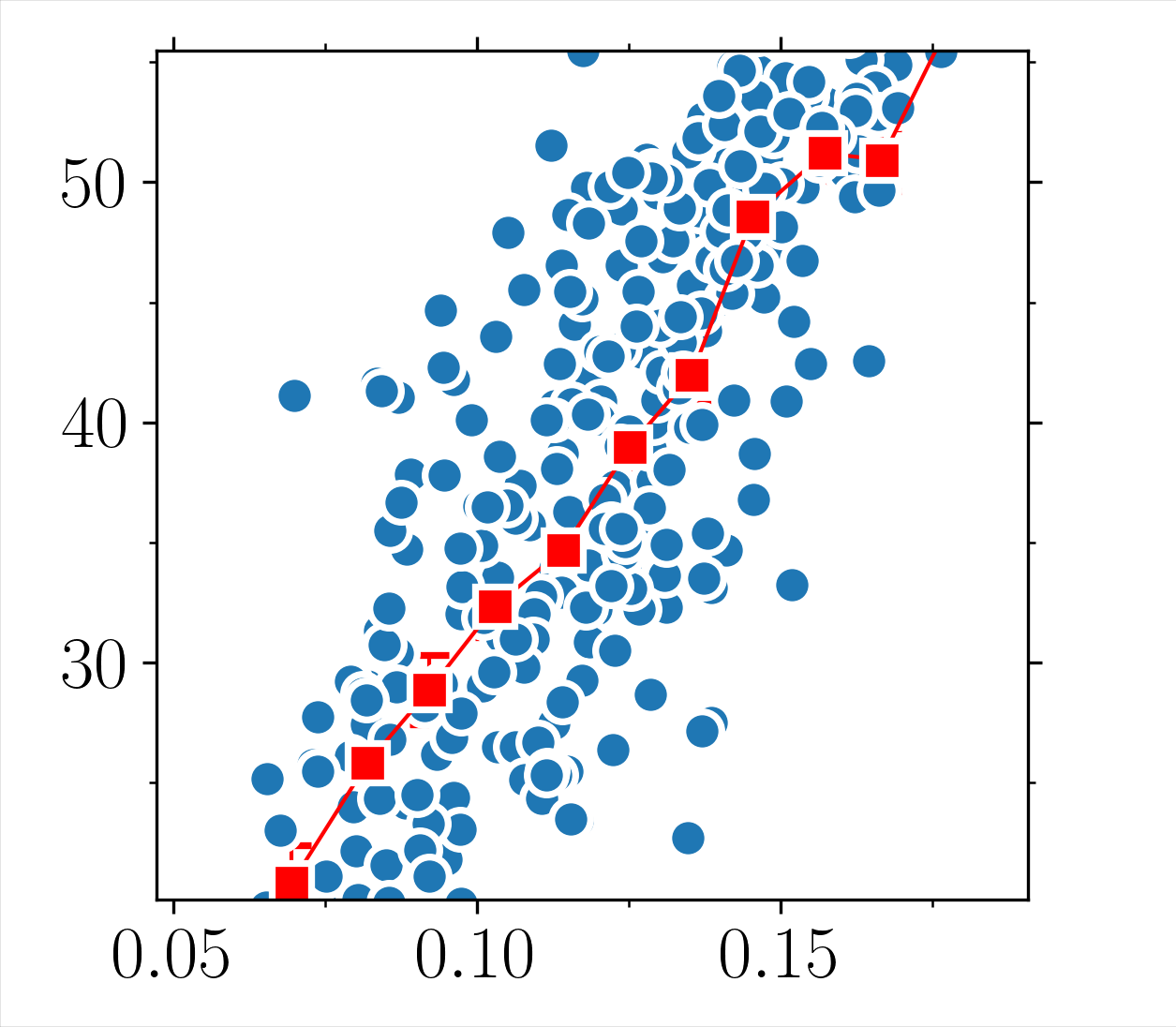}
        \LabelFig{16}{87}{$c)$ \scriptsize \gltwo}
        \Labelxy{50}{0}{0}{$p$}
        \Labelxy{-8}{46}{90}{{$\xi$\tiny (\r{A})}}
    \end{overpic}
    %

    \begin{overpic}[width=0.23\textwidth]{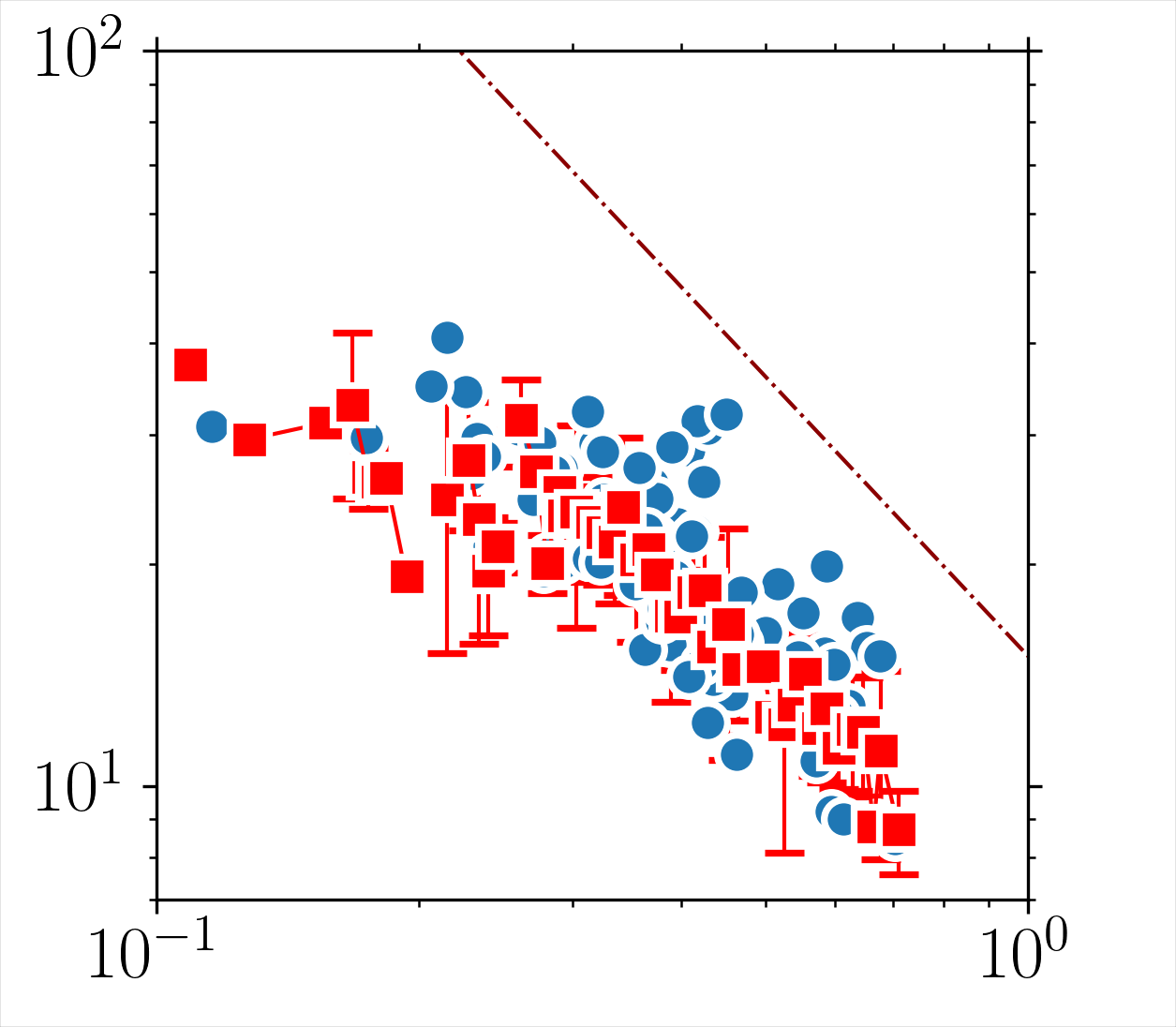}
        \LabelFig{16}{76}{$b)$}
        \Labelxy{-8}{46}{90}{{$\xi$\tiny (\r{A})}}
        \Labelxy{38}{-3}{0}{$1-p/p_\text{max}$}
        \begin{tikzpicture}
          \coordinate (a) at (0,0);
            \node[white] at (a) {\tiny.};               %
            \drawSlope{2.1}{2.8}{0.25}{225}{darkred}{$\hspace{-5pt}\scriptstyle\nu$}
            \legLine{0.7}{0.6}{darkred}{$\scriptstyle \xi\propto (p_{\text{max}}-p)^{-\nu}$}{0.9}
		\end{tikzpicture}
    \end{overpic}
    \begin{overpic}[width=0.23\textwidth]{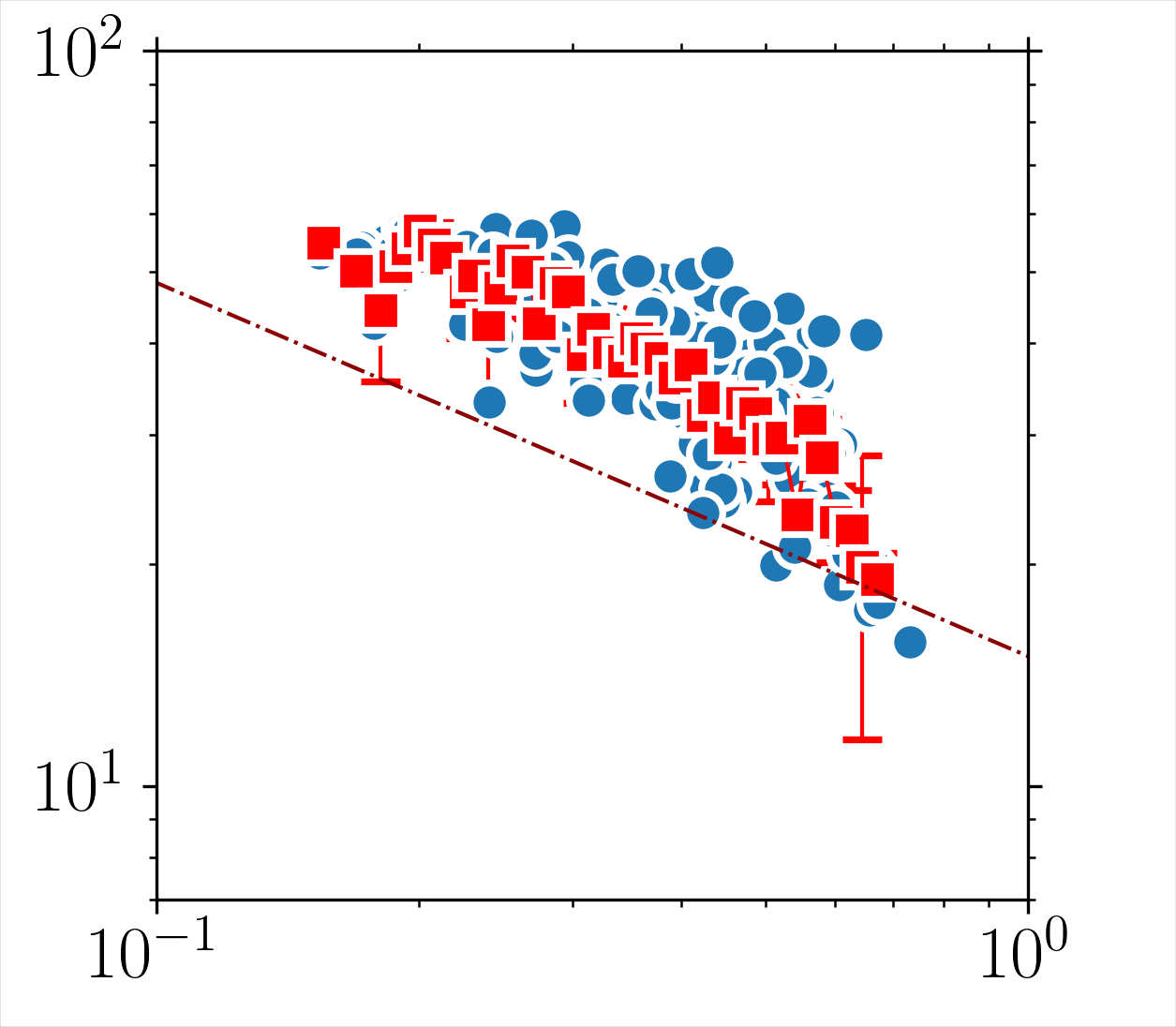}
        \LabelFig{16}{76}{$d)$}
        \Labelxy{-8}{46}{90}{{$\xi$\tiny (\r{A})}}
        \Labelxy{38}{-3}{0}{$1-p/p_\text{max}$}
        \begin{tikzpicture}
          \coordinate (a) at (0,0);
            \node[white] at (a) {\tiny.};               %
            \drawSlope{1.8}{1.8}{0.25}{64}{darkred}{$\scriptstyle\nu$}
            \legLine{0.7}{0.6}{darkred}{$\scriptstyle \xi\propto (p_{\text{max}}-p)^{-\nu}$}{0.9}
		\end{tikzpicture}
    \end{overpic}
    \caption{Correlation length $\xi$ plotted against $p$ corresponding to the \textbf{a}) \glone and \textbf{c}) \gltwo glasses. The graphs in \textbf{b}) and \textbf{d}) are the same as \textbf{a}) and \textbf{c}) but plot $\xi$ as a function of $1-p/p_\text{max}$ with \textbf{a}) $p_{\text{max}}= 0.14$ and \textbf{c}) $p_{\text{max}}= 0.20$. The (red) curves indicate binning averaged data. The dashdotted lines denote power laws $\xi\propto (p_{\text{max}}-p)^{-\nu}$ with \textbf{b}) $\nu=1.26$ and \textbf{d}) $\nu=0.51$.}
    \label{fig:crltn}
\end{figure}
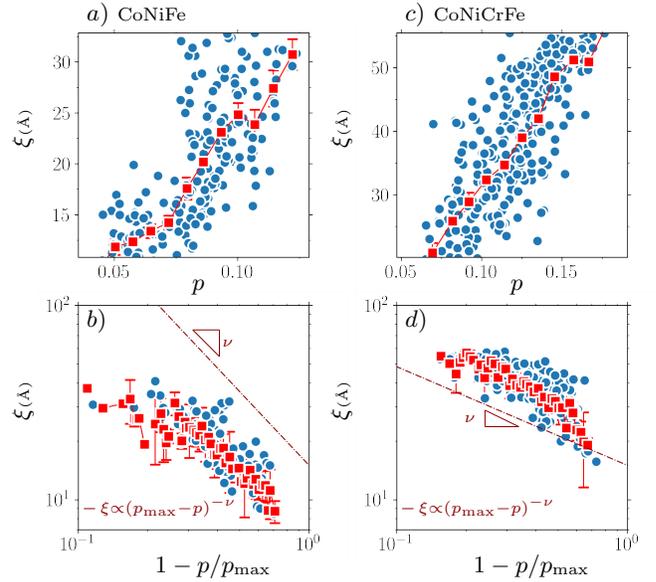

\begingroup
\begin{center}

\begin{table*}[t]
\begin{tabular}{|p{3.0cm}|p{2.2cm}|p{2.4cm}|p{2.4cm}|p{2.2cm}} 
\hline\hline 
\begin{tabular}{@{}l@{}}\textbf{Chemical} \\ \textbf{Composition}  \end{tabular} &
\begin{tabular}{@{}l@{}}\textbf{Cluster Size} \\ \textbf{Distribution} \\ $n_s \propto s^{-\tau}$ \\ $\tau$  \end{tabular} &
\begin{tabular}{@{}l@{}}\textbf{Mean Cluster } \\ \textbf{Size} \\ $S\propto (p-p_\text{max})^{-\gamma}$ \\ $\gamma$ \end{tabular} &
\begin{tabular}{@{}l@{}}\small\textbf{Correlation} \\ \textbf{Length} \\ $\xi\propto (p-p_\text{max})^{-\nu}$ \\ $\nu$ \end{tabular} &
\begin{tabular}{@{}l@{}}\small\textbf{Fractal} \\ \textbf{Dimension} \\ $s\propto r_s^{d_f}$ \\ $d_f$ \end{tabular} \\[2ex]
\hline\hline  
\glzero  & $1.96$ & $2.23$ & $0.87$ & $2.27$  \\[1ex] 
\glone   & $1.98$ & $2.89$ & $1.26$ & $2.25$ \\[1ex]
\gltwo   & $2.05$ & $2.20$ & $0.51$ & $2.62$ \\[1ex]
\glthree & $2.02$ & $2.42$ & $0.95$ & $2.48$ \\[1ex]
\glfour  & $2.00$ & $2.26$ & $0.62$ & $2.43$ \\[1ex]
\glfive  & $1.93$ & $2.69$ & $0.94$ & $2.28$ \\[1ex]
CuZr\cite{ghosh2017direct}  & $-$ & $-$ & $0.85\pm 0.1$ & $2.0$ \\[1ex]
\begin{tabular}{@{}l@{}} Percolation \\ ($\emph{d}=3$) \cite{stauffer2018introduction} \end{tabular}  & $2.18$ & $1.80$ & $0.88$ & $2.53$ \\[2ex]
\begin{tabular}{@{}l@{}} Directed Percolation\\  ($\emph{d}=3$) \cite{hinrichsen2000non} \end{tabular}  & $-$ & $1.25$ & $0.6$ & $2.0$ \\

\hline 
\end{tabular}
\caption{Comparison between estimated scaling exponents associated with different chemical compositions and three dimensional ($d=3$) percolation theory.}
\label{table:exponents}
\end{table*}
\end{center}
\endgroup

\begin{figure}[b]
    \centering
    \begin{overpic}[width=0.23\textwidth]{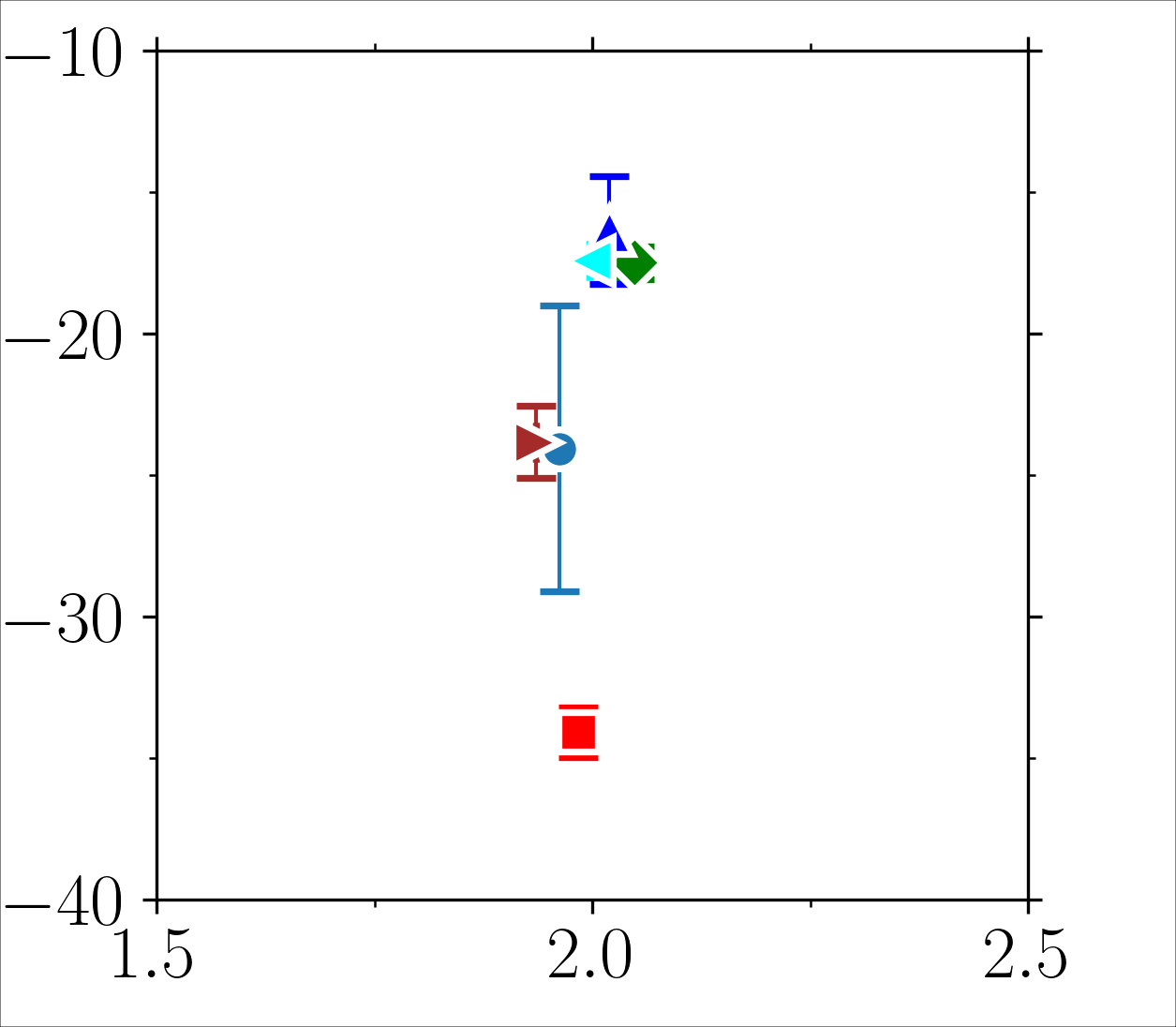}
        \LabelFig{15}{76}{$a)$}
        \Labelxy{50}{-3}{0}{$\tau$}
        \Labelxy{-10}{26}{90}{$h_\text{min}$\tiny(Gpa)}
    \end{overpic}
   \begin{overpic}[width=0.23\textwidth]{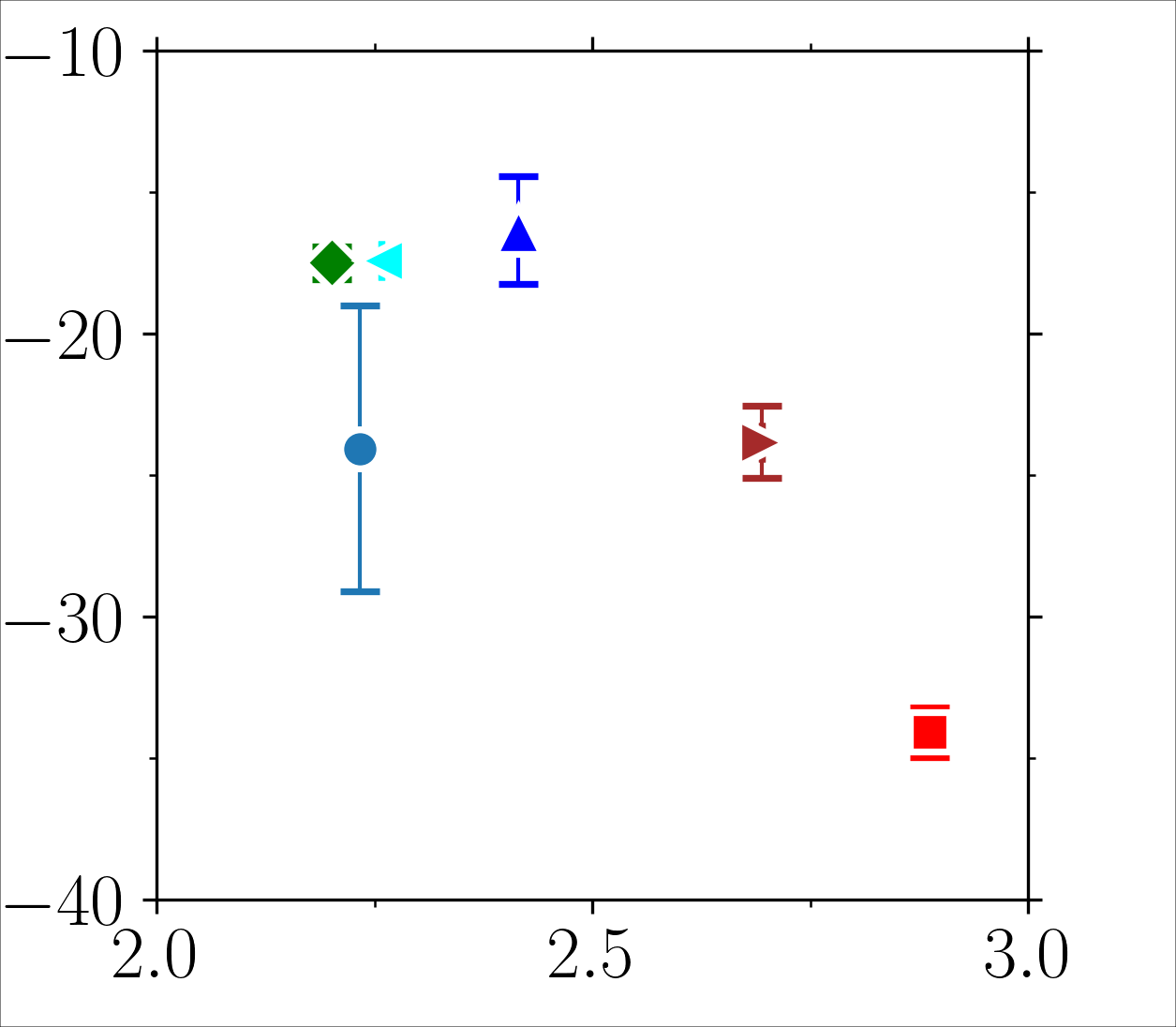}
        \LabelFig{15}{76}{$b)$}
        \Labelxy{50}{-3}{0}{$\gamma$}
    \end{overpic}
    \begin{overpic}[width=0.23\textwidth]{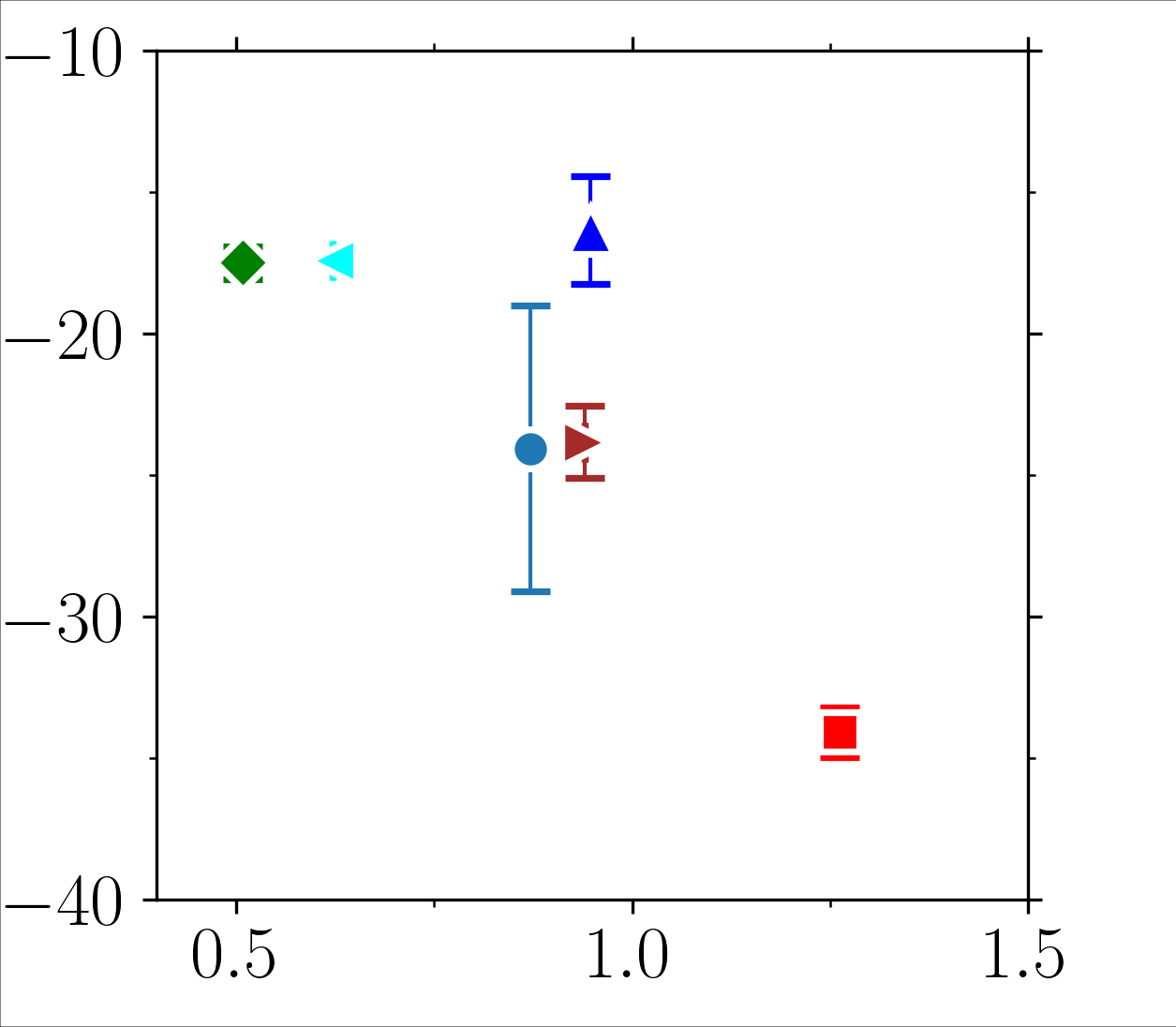}
        \LabelFig{15}{76}{$c)$}
        \Labelxy{50}{-3}{0}{$\nu$}
        \Labelxy{-10}{26}{90}{$h_\text{min}$\tiny(Gpa)}
    \end{overpic}
    \begin{overpic}[width=0.23\textwidth]{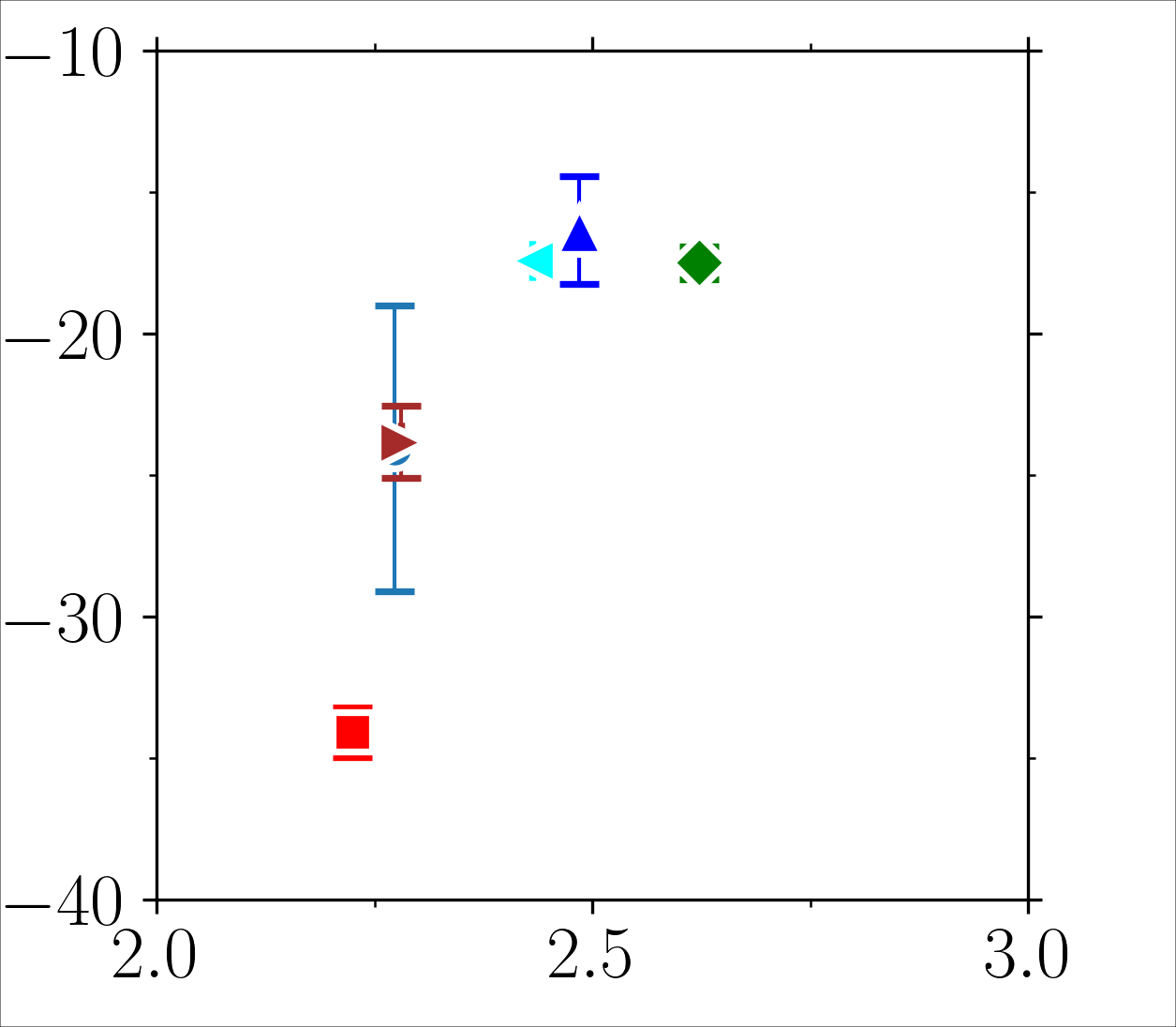}
        \LabelFig{15}{76}{$c)$}
        \Labelxy{50}{-3}{0}{$d_f$}
    \end{overpic}
    %
    %
    \caption{Scatter plot of softening modulus $h_\text{min}$ and scaling exponents \remove{\textbf{a}) $\beta$} \textbf{a}) $\gamma$ \textbf{b}) $\nu$ corresponding to the \glzero(\protect\circTxtFill{0}{0}{blue}), \glone(\protect\legSqTxt{0}{0}{red}), \gltwo(\protect\legDiamondTxt{0}{0}{darkspringgreen}), \glthree(\protect\legTriangleTxt{0}{0}{blue}), \glfour(\protect\leftTriang{0}{0}{cyan}), and \glfive (\protect\rightTriang{0}{0}{brown}) bulk metallic glasses.}
    \label{fig:hmin_gamma}
\end{figure}

The above elasticity maps give a visual impression that the glass failure (at $\gmax \simeq 0.1$) might indeed coincide with a percolation transition of \add{softness at \pmax} upon shear loading. 
In order to validate this picture, we adopted ideas from the classical percolation theory \cite{stauffer2018introduction} including investigations of cluster sizes and their dynamical evolution. 
As a basic statistical property, $n_s$ denotes the probability distribution function associated with the number of clusters containing $s$ unstable sites.
Figure~\ref{fig:clusterSize}(a) and (c) plots $n_s$ associated with the \glone and \gltwo glasses at three different strains. 
The cluster size distributions tend to develop fairly long tails as $p$ approaches \pmax.
Our data suggest a robust power-law decay $n_s\propto s^{-\tau}$ with $\tau= 2.0$ and $2.1$, associated with \glone and \gltwo, over at least two decades in $s$. 
The estimated range of exponent $\tau$ is fairly robust showing minimal compositional dependence (as reported in Table~\ref{table:exponents}) and/or variations near \pmax.
The radius of gyration associated with a cluster of size $s$ may be defined as $r^2_s=\sum_{i=1}^{s}|\vec{r}_i-\vec{r}_0|^2/s$ with the center of mass $\vec{r}_0=\sum_{i=1}^{s}\vec{r_i}/s$.
Figure~\ref{fig:clusterSize}(b) and (d) illustrates that $s\propto r_s^{d_f}$ with fractal dimension $d_f= 2.3$, $2.6$ corresponding to \glone and \gltwo, respectively.  
This almost agrees with Fig.~\ref{fig:loadCurve}(b) and (c) in that the soft spots tend to form fairly compact clusters in the latter glass whereas the former one is associated with more localized (but still system-spanning) features.   

Figure~\ref{fig:smean}(a) and (c) displays the mean cluster size $S={\sum_s n_ss^2}/{\sum_s n_ss}$ and its evolution with $p$.
The average size reveals a certain algebraic divergence of $S$ on approach to the maximum fraction $p\rightarrow p_\text{max}$.
This divergence is demonstrated in Fig.~\ref{fig:smean}(b) and (d) with the mean size scaling as $S\propto (p_\text{max}-p)^{-\gamma}$ and $\gamma= 2.89$, $2.20$ corresponding to the \glone and \gltwo compositions, respectively.
The proposed scaling is valid for a little less than a decade in $1-p/p_\text{max}$ down to a roll-off at small arguments potentially due to finite size effects.
The (squared) correlation length $\xi^2={2\sum_s r^2_ss^2n_s}/{\sum_ss^2n_s}$ is defined based on a weighted average associated with the radius of gyration $r^2_s=\sum_{i=1}^{s}|\vec{r}_i-\vec{r}_0|^2/s$ of a cluster of size $s$, as shown in Fig.~\ref{fig:crltn}(a) and (c).
Here the center of mass is $\vec{r}_0=\sum_{i=1}^{s}\vec{r_i}/s$.
As $p$ increase toward \pmax in Fig.~\ref{fig:crltn}(b) and (d), the correlation length scales like $\xi\propto (p_{\text{max}}-p)^{-\nu}$ with $\nu= 1.26$ corresponding to \glone and $\nu=0.51$ for \gltwo.
The center and right panels of Fig.~\ref{fig:statistics_gamma} show  the mean cluster size $S$ and correlation length $\xi$ as a function of applied strain $\gamma_{xy}$ associated with different compositions.

We repeated the above analysis to infer critical exponents corresponding to the other chemical compositions (see Fig.~\ref{fig:statistics_p} and Table~\ref{table:exponents}) and sought for potential connections with the \change{glass post-yielding (softening) behavior}{softening modulus \hmin}.  
Figure~\ref{fig:hmin_gamma} plots the latter and estimated critical exponents $\tau$, $\gamma$, $\nu$, and $d_f$ for the \glzero, \glone, \gltwo, \glthree, \glfour, \glfive metallic glasses.
As previously discussed, the cluster size exponent $\tau$ in Fig.~\ref{fig:hmin_gamma}(a) shows insignificant variations with chemical compositions.
The scatter plot of \hmin and mean cluster size exponent $\gamma$ in Fig.~\ref{fig:hmin_gamma}(b) suggests noticeable anti-correlations between the two observables.
\remove{This is a remarkable observation suggesting that a steeper evolution of elastic inhomogeneity near failure transition leads to a better glass (lower \hmin).}
\remove{Likewise, the percolation probability exponent $\beta$ in Fig.~\ref{fig:hmin_gamma}(b) reveals a non-trivial scaling with \hmin.
Positive correlations in this case may imply that larger susceptibilities (and therefore lower $\beta$ exponents) should, on average, result in a sharper yielding transition.
We remark that only the \glone glass gives $\beta < 1$ as a commonly-observed scaling for order parameters characterizing second-order phase transitions (with a diverging susceptibility) \cite{?}.}
\change{The correlation length exponent $\nu$, on the other hand, seems to be statistically uncorrelated with \hmin as in Fig.~\ref{fig:hmin_gamma}(c).}{Likewise, the correlation length exponent $\nu$ seems to be statistically (anti-)correlated with \hmin as in Fig.~\ref{fig:hmin_gamma}(c).}
The increasing trend in the fractal dimension $d_f$ might be indicative of localized (system-spanning) soft spots in \emph{good} glasses that become increasingly packed in space as compositional variations lead to a smoother yielding transition.
Given that there are only two independent exponents in standard percolation theory, the behavior of $\nu$ and $\gamma$ should be largely controlled by variabilities in $d_f$.

\emph{Conclusions \& Discussions---}
We have presented direct evidence that \change{ductility}{yielding transition} of bulk metallic glasses is \remove{strongly dependent on how chemical compositions fine tune}{is accompanied with} percolation threshold of softness upon failure across different chemical compositions. 
The former has been quantified by the softening modulus \hmin, indicative of the sharpness of plastic yielding transition, and the latter was characterized by analyzing connected networks of mechanically soft regions that grow under application of external stress.
Our analysis further indicates critical scaling features associated with the dynamics and topology of soft clusters suggesting a close relevance of the percolation transition.
Relevant scaling exponents and potential correlations with the macroscopic stress-strain curve have been considered from the compositional-dependence perspective.
Table~\ref{table:exponents} features the range of estimated exponents corresponding to different multi-component metallic glasses probed in this work and the study by Schall et al. \cite{ghosh2017direct} as well as those inferred from the percolation theory.
In this framework, we find a fairly robust range (across different systems) associated with the cluster size exponent $1.9\le\tau\le2.1$ and comparable with theoretical predictions \cite{stauffer2018introduction}. 
\remove{Likewise, the divergence of correlation length and associated exponent $0.5\le\nu\le1.3$ does not show an appreciable composition-dependence.}
On the other hand, $\gamma$, $\nu$, and $d_f$, inferred from the scaling of the correlation lengths and mean cluster sizes, somewhat correlate with softening modulus \hmin suggesting that the softness percolation might influence macroscopic yielding properties through compositional dependence.
Such correlations may also suggest that the notion of \emph{universality} in deformation and flow properties of multicomponent glasses could be limited owing to compositional/microstructural associations.
Nevertheless, given a relatively narrow range of the softness exponents across different glasses, one might still consider the same universality class for the observed transition despite the fact that our results show substantial composition-based variabilities in yielding properties.


\begin{acknowledgments}
This research was funded by the European Union Horizon 2020 research and innovation program under grant agreement no. 857470 and from the European Regional Development Fund via Foundation for Polish Science International Research Agenda PLUS program grant no. MAB PLUS/2018/8.
We wish to thank A. Esfandiarpour and R. Alvarez for providing the data sets.
\end{acknowledgments}

\bibliography{references}

\begin{thebibliography}{34}%
\makeatletter
\providecommand \@ifxundefined [1]{%
 \@ifx{#1\undefined}
}%
\providecommand \@ifnum [1]{%
 \ifnum #1\expandafter \@firstoftwo
 \else \expandafter \@secondoftwo
 \fi
}%
\providecommand \@ifx [1]{%
 \ifx #1\expandafter \@firstoftwo
 \else \expandafter \@secondoftwo
 \fi
}%
\providecommand \natexlab [1]{#1}%
\providecommand \enquote  [1]{``#1''}%
\providecommand \bibnamefont  [1]{#1}%
\providecommand \bibfnamefont [1]{#1}%
\providecommand \citenamefont [1]{#1}%
\providecommand \href@noop [0]{\@secondoftwo}%
\providecommand \href [0]{\begingroup \@sanitize@url \@href}%
\providecommand \@href[1]{\@@startlink{#1}\@@href}%
\providecommand \@@href[1]{\endgroup#1\@@endlink}%
\providecommand \@sanitize@url [0]{\catcode `\\12\catcode `\$12\catcode
  `\&12\catcode `\#12\catcode `\^12\catcode `\_12\catcode `\%12\relax}%
\providecommand \@@startlink[1]{}%
\providecommand \@@endlink[0]{}%
\providecommand \url  [0]{\begingroup\@sanitize@url \@url }%
\providecommand \@url [1]{\endgroup\@href {#1}{\urlprefix }}%
\providecommand \urlprefix  [0]{URL }%
\providecommand \Eprint [0]{\href }%
\providecommand \doibase [0]{https://doi.org/}%
\providecommand \selectlanguage [0]{\@gobble}%
\providecommand \bibinfo  [0]{\@secondoftwo}%
\providecommand \bibfield  [0]{\@secondoftwo}%
\providecommand \translation [1]{[#1]}%
\providecommand \BibitemOpen [0]{}%
\providecommand \bibitemStop [0]{}%
\providecommand \bibitemNoStop [0]{.\EOS\space}%
\providecommand \EOS [0]{\spacefactor3000\relax}%
\providecommand \BibitemShut  [1]{\csname bibitem#1\endcsname}%
\let\auto@bib@innerbib\@empty
\bibitem [{\citenamefont {Dai}\ and\ \citenamefont {Bai}(2008)}]{dai2008basic}%
  \BibitemOpen
  \bibfield  {author} {\bibinfo {author} {\bibfnamefont {L.}~\bibnamefont
  {Dai}}\ and\ \bibinfo {author} {\bibfnamefont {Y.}~\bibnamefont {Bai}},\
  }\bibfield  {title} {\bibinfo {title} {Basic mechanical behaviors and
  mechanics of shear banding in bmgs},\ }\href@noop {} {\bibfield  {journal}
  {\bibinfo  {journal} {International Journal of Impact Engineering}\ }\textbf
  {\bibinfo {volume} {35}},\ \bibinfo {pages} {704} (\bibinfo {year}
  {2008})}\BibitemShut {NoStop}%
\bibitem [{\citenamefont {Schuh}\ \emph {et~al.}(2007)\citenamefont {Schuh},
  \citenamefont {Hufnagel},\ and\ \citenamefont
  {Ramamurty}}]{schuh2007mechanical}%
  \BibitemOpen
  \bibfield  {author} {\bibinfo {author} {\bibfnamefont {C.~A.}\ \bibnamefont
  {Schuh}}, \bibinfo {author} {\bibfnamefont {T.~C.}\ \bibnamefont
  {Hufnagel}},\ and\ \bibinfo {author} {\bibfnamefont {U.}~\bibnamefont
  {Ramamurty}},\ }\bibfield  {title} {\bibinfo {title} {Mechanical behavior of
  amorphous alloys},\ }\href@noop {} {\bibfield  {journal} {\bibinfo  {journal}
  {Acta Materialia}\ }\textbf {\bibinfo {volume} {55}},\ \bibinfo {pages}
  {4067} (\bibinfo {year} {2007})}\BibitemShut {NoStop}%
\bibitem [{\citenamefont {Ding}\ and\ \citenamefont
  {Yao}(2013)}]{ding2013high}%
  \BibitemOpen
  \bibfield  {author} {\bibinfo {author} {\bibfnamefont {H.}~\bibnamefont
  {Ding}}\ and\ \bibinfo {author} {\bibfnamefont {K.}~\bibnamefont {Yao}},\
  }\bibfield  {title} {\bibinfo {title} {High entropy ti20zr20cu20ni20be20 bulk
  metallic glass},\ }\href@noop {} {\bibfield  {journal} {\bibinfo  {journal}
  {Journal of non-crystalline solids}\ }\textbf {\bibinfo {volume} {364}},\
  \bibinfo {pages} {9} (\bibinfo {year} {2013})}\BibitemShut {NoStop}%
\bibitem [{\citenamefont {Karimi}\ \emph {et~al.}(2022)\citenamefont {Karimi},
  \citenamefont {Esfandiarpour}, \citenamefont {Alvarez-Donado}, \citenamefont
  {Alava},\ and\ \citenamefont {Papanikolaou}}]{karimi2021shear}%
  \BibitemOpen
  \bibfield  {author} {\bibinfo {author} {\bibfnamefont {K.}~\bibnamefont
  {Karimi}}, \bibinfo {author} {\bibfnamefont {A.}~\bibnamefont
  {Esfandiarpour}}, \bibinfo {author} {\bibfnamefont {R.}~\bibnamefont
  {Alvarez-Donado}}, \bibinfo {author} {\bibfnamefont {M.~J.}\ \bibnamefont
  {Alava}},\ and\ \bibinfo {author} {\bibfnamefont {S.}~\bibnamefont
  {Papanikolaou}},\ }\bibfield  {title} {\bibinfo {title} {Shear banding
  instability in multicomponent metallic glasses: Interplay of composition and
  short-range order},\ }\href@noop {} {\bibfield  {journal} {\bibinfo
  {journal} {Physical Review B}\ }\textbf {\bibinfo {volume} {105}},\ \bibinfo
  {pages} {094117} (\bibinfo {year} {2022})}\BibitemShut {NoStop}%
\bibitem [{\citenamefont {Denisov}\ \emph {et~al.}(2017)\citenamefont
  {Denisov}, \citenamefont {L{\H{o}}rincz}, \citenamefont {Wright},
  \citenamefont {Hufnagel}, \citenamefont {Nawano}, \citenamefont {Gu},
  \citenamefont {Uhl}, \citenamefont {Dahmen},\ and\ \citenamefont
  {Schall}}]{denisov2017universal}%
  \BibitemOpen
  \bibfield  {author} {\bibinfo {author} {\bibfnamefont {D.~V.}\ \bibnamefont
  {Denisov}}, \bibinfo {author} {\bibfnamefont {K.~A.}\ \bibnamefont
  {L{\H{o}}rincz}}, \bibinfo {author} {\bibfnamefont {W.~J.}\ \bibnamefont
  {Wright}}, \bibinfo {author} {\bibfnamefont {T.~C.}\ \bibnamefont
  {Hufnagel}}, \bibinfo {author} {\bibfnamefont {A.}~\bibnamefont {Nawano}},
  \bibinfo {author} {\bibfnamefont {X.}~\bibnamefont {Gu}}, \bibinfo {author}
  {\bibfnamefont {J.~T.}\ \bibnamefont {Uhl}}, \bibinfo {author} {\bibfnamefont
  {K.~A.}\ \bibnamefont {Dahmen}},\ and\ \bibinfo {author} {\bibfnamefont
  {P.}~\bibnamefont {Schall}},\ }\bibfield  {title} {\bibinfo {title}
  {Universal slip dynamics in metallic glasses and granular matter--linking
  frictional weakening with inertial effects},\ }\href@noop {} {\bibfield
  {journal} {\bibinfo  {journal} {Scientific reports}\ }\textbf {\bibinfo
  {volume} {7}},\ \bibinfo {pages} {1} (\bibinfo {year} {2017})}\BibitemShut
  {NoStop}%
\bibitem [{\citenamefont {Cheng}\ and\ \citenamefont
  {Ma}(2011)}]{cheng2011atomic}%
  \BibitemOpen
  \bibfield  {author} {\bibinfo {author} {\bibfnamefont {Y.}~\bibnamefont
  {Cheng}}\ and\ \bibinfo {author} {\bibfnamefont {E.}~\bibnamefont {Ma}},\
  }\bibfield  {title} {\bibinfo {title} {Atomic-level structure and
  structure--property relationship in metallic glasses},\ }\href@noop {}
  {\bibfield  {journal} {\bibinfo  {journal} {Progress in materials science}\
  }\textbf {\bibinfo {volume} {56}},\ \bibinfo {pages} {379} (\bibinfo {year}
  {2011})}\BibitemShut {NoStop}%
\bibitem [{\citenamefont {Cheng}\ \emph {et~al.}(2009)\citenamefont {Cheng},
  \citenamefont {Cao},\ and\ \citenamefont {Ma}}]{cheng2009correlation}%
  \BibitemOpen
  \bibfield  {author} {\bibinfo {author} {\bibfnamefont {Y.}~\bibnamefont
  {Cheng}}, \bibinfo {author} {\bibfnamefont {A.}~\bibnamefont {Cao}},\ and\
  \bibinfo {author} {\bibfnamefont {E.}~\bibnamefont {Ma}},\ }\bibfield
  {title} {\bibinfo {title} {Correlation between the elastic modulus and the
  intrinsic plastic behavior of metallic glasses: The roles of atomic
  configuration and alloy composition},\ }\href@noop {} {\bibfield  {journal}
  {\bibinfo  {journal} {Acta Materialia}\ }\textbf {\bibinfo {volume} {57}},\
  \bibinfo {pages} {3253} (\bibinfo {year} {2009})}\BibitemShut {NoStop}%
\bibitem [{\citenamefont {Wang}\ \emph {et~al.}(2018)\citenamefont {Wang},
  \citenamefont {Ding}, \citenamefont {Yan}, \citenamefont {Asta},
  \citenamefont {Ritchie},\ and\ \citenamefont {Li}}]{wang2018spatial}%
  \BibitemOpen
  \bibfield  {author} {\bibinfo {author} {\bibfnamefont {N.}~\bibnamefont
  {Wang}}, \bibinfo {author} {\bibfnamefont {J.}~\bibnamefont {Ding}}, \bibinfo
  {author} {\bibfnamefont {F.}~\bibnamefont {Yan}}, \bibinfo {author}
  {\bibfnamefont {M.}~\bibnamefont {Asta}}, \bibinfo {author} {\bibfnamefont
  {R.~O.}\ \bibnamefont {Ritchie}},\ and\ \bibinfo {author} {\bibfnamefont
  {L.}~\bibnamefont {Li}},\ }\bibfield  {title} {\bibinfo {title} {Spatial
  correlation of elastic heterogeneity tunes the deformation behavior of
  metallic glasses},\ }\href@noop {} {\bibfield  {journal} {\bibinfo  {journal}
  {npj Computational Materials}\ }\textbf {\bibinfo {volume} {4}},\ \bibinfo
  {pages} {1} (\bibinfo {year} {2018})}\BibitemShut {NoStop}%
\bibitem [{\citenamefont {Wu}\ \emph {et~al.}(2009)\citenamefont {Wu},
  \citenamefont {Zhang},\ and\ \citenamefont {Mao}}]{wu2009transition}%
  \BibitemOpen
  \bibfield  {author} {\bibinfo {author} {\bibfnamefont {F.-F.}\ \bibnamefont
  {Wu}}, \bibinfo {author} {\bibfnamefont {Z.-F.}\ \bibnamefont {Zhang}},\ and\
  \bibinfo {author} {\bibfnamefont {S.~X.-Y.}\ \bibnamefont {Mao}},\ }\bibfield
   {title} {\bibinfo {title} {Transition of failure mode and enhanced plastic
  deformation of metallic glass by multiaxial confinement},\ }\href@noop {}
  {\bibfield  {journal} {\bibinfo  {journal} {Advanced Engineering Materials}\
  }\textbf {\bibinfo {volume} {11}},\ \bibinfo {pages} {898} (\bibinfo {year}
  {2009})}\BibitemShut {NoStop}%
\bibitem [{\citenamefont {Chen}\ \emph {et~al.}(2006)\citenamefont {Chen},
  \citenamefont {Inoue}, \citenamefont {Zhang},\ and\ \citenamefont
  {Sakurai}}]{chen2006extraordinary}%
  \BibitemOpen
  \bibfield  {author} {\bibinfo {author} {\bibfnamefont {M.}~\bibnamefont
  {Chen}}, \bibinfo {author} {\bibfnamefont {A.}~\bibnamefont {Inoue}},
  \bibinfo {author} {\bibfnamefont {W.}~\bibnamefont {Zhang}},\ and\ \bibinfo
  {author} {\bibfnamefont {T.}~\bibnamefont {Sakurai}},\ }\bibfield  {title}
  {\bibinfo {title} {Extraordinary plasticity of ductile bulk metallic
  glasses},\ }\href@noop {} {\bibfield  {journal} {\bibinfo  {journal}
  {Physical Review Letters}\ }\textbf {\bibinfo {volume} {96}},\ \bibinfo
  {pages} {245502} (\bibinfo {year} {2006})}\BibitemShut {NoStop}%
\bibitem [{\citenamefont {Cheng}\ \emph {et~al.}(2008)\citenamefont {Cheng},
  \citenamefont {Cao}, \citenamefont {Sheng},\ and\ \citenamefont
  {Ma}}]{cheng2008local}%
  \BibitemOpen
  \bibfield  {author} {\bibinfo {author} {\bibfnamefont {Y.}~\bibnamefont
  {Cheng}}, \bibinfo {author} {\bibfnamefont {A.~J.}\ \bibnamefont {Cao}},
  \bibinfo {author} {\bibfnamefont {H.}~\bibnamefont {Sheng}},\ and\ \bibinfo
  {author} {\bibfnamefont {E.}~\bibnamefont {Ma}},\ }\bibfield  {title}
  {\bibinfo {title} {Local order influences initiation of plastic flow in
  metallic glass: Effects of alloy composition and sample cooling history},\
  }\href@noop {} {\bibfield  {journal} {\bibinfo  {journal} {Acta Materialia}\
  }\textbf {\bibinfo {volume} {56}},\ \bibinfo {pages} {5263} (\bibinfo {year}
  {2008})}\BibitemShut {NoStop}%
\bibitem [{\citenamefont {Shi}\ and\ \citenamefont
  {Falk}(2005)}]{shi2005strain}%
  \BibitemOpen
  \bibfield  {author} {\bibinfo {author} {\bibfnamefont {Y.}~\bibnamefont
  {Shi}}\ and\ \bibinfo {author} {\bibfnamefont {M.~L.}\ \bibnamefont {Falk}},\
  }\bibfield  {title} {\bibinfo {title} {Strain localization and percolation of
  stable structure in amorphous solids},\ }\href@noop {} {\bibfield  {journal}
  {\bibinfo  {journal} {Physical review letters}\ }\textbf {\bibinfo {volume}
  {95}},\ \bibinfo {pages} {095502} (\bibinfo {year} {2005})}\BibitemShut
  {NoStop}%
\bibitem [{\citenamefont {Albano}\ and\ \citenamefont
  {Falk}(2005)}]{albano2005shear}%
  \BibitemOpen
  \bibfield  {author} {\bibinfo {author} {\bibfnamefont {F.}~\bibnamefont
  {Albano}}\ and\ \bibinfo {author} {\bibfnamefont {M.~L.}\ \bibnamefont
  {Falk}},\ }\bibfield  {title} {\bibinfo {title} {Shear softening and
  structure in a simulated three-dimensional binary glass},\ }\href@noop {}
  {\bibfield  {journal} {\bibinfo  {journal} {The Journal of chemical physics}\
  }\textbf {\bibinfo {volume} {122}},\ \bibinfo {pages} {154508} (\bibinfo
  {year} {2005})}\BibitemShut {NoStop}%
\bibitem [{\citenamefont {Wang}(2018)}]{wang2018multi}%
  \BibitemOpen
  \bibfield  {author} {\bibinfo {author} {\bibfnamefont {N.}~\bibnamefont
  {Wang}},\ }\emph {\bibinfo {title} {Multi-scale modeling of spatial
  heterogeneity effect on the shear banding behaviors in metallic glasses}},\
  \href@noop {} {Ph.D. thesis},\ \bibinfo  {school} {The University of Alabama}
  (\bibinfo {year} {2018})\BibitemShut {NoStop}%
\bibitem [{\citenamefont {Nagamanasa}\ \emph {et~al.}(2014)\citenamefont
  {Nagamanasa}, \citenamefont {Gokhale}, \citenamefont {Sood},\ and\
  \citenamefont {Ganapathy}}]{nagamanasa2014direct}%
  \BibitemOpen
  \bibfield  {author} {\bibinfo {author} {\bibfnamefont {K.~H.}\ \bibnamefont
  {Nagamanasa}}, \bibinfo {author} {\bibfnamefont {S.}~\bibnamefont {Gokhale}},
  \bibinfo {author} {\bibfnamefont {A.}~\bibnamefont {Sood}},\ and\ \bibinfo
  {author} {\bibfnamefont {R.}~\bibnamefont {Ganapathy}},\ }\bibfield  {title}
  {\bibinfo {title} {Direct evidence for an absorbing phase transition
  governing yielding of a soft glass},\ }\href@noop {} {\bibfield  {journal}
  {\bibinfo  {journal} {arXiv preprint arXiv:1402.2730}\ } (\bibinfo {year}
  {2014})}\BibitemShut {NoStop}%
\bibitem [{\citenamefont {Shrivastav}\ \emph {et~al.}(2016)\citenamefont
  {Shrivastav}, \citenamefont {Chaudhuri},\ and\ \citenamefont
  {Horbach}}]{shrivastav2016yielding}%
  \BibitemOpen
  \bibfield  {author} {\bibinfo {author} {\bibfnamefont {G.~P.}\ \bibnamefont
  {Shrivastav}}, \bibinfo {author} {\bibfnamefont {P.}~\bibnamefont
  {Chaudhuri}},\ and\ \bibinfo {author} {\bibfnamefont {J.}~\bibnamefont
  {Horbach}},\ }\bibfield  {title} {\bibinfo {title} {Yielding of glass under
  shear: A directed percolation transition precedes shear-band formation},\
  }\href@noop {} {\bibfield  {journal} {\bibinfo  {journal} {Physical Review
  E}\ }\textbf {\bibinfo {volume} {94}},\ \bibinfo {pages} {042605} (\bibinfo
  {year} {2016})}\BibitemShut {NoStop}%
\bibitem [{\citenamefont {Ghosh}\ \emph {et~al.}(2017)\citenamefont {Ghosh},
  \citenamefont {Budrikis}, \citenamefont {Chikkadi}, \citenamefont {Sellerio},
  \citenamefont {Zapperi},\ and\ \citenamefont {Schall}}]{ghosh2017direct}%
  \BibitemOpen
  \bibfield  {author} {\bibinfo {author} {\bibfnamefont {A.}~\bibnamefont
  {Ghosh}}, \bibinfo {author} {\bibfnamefont {Z.}~\bibnamefont {Budrikis}},
  \bibinfo {author} {\bibfnamefont {V.}~\bibnamefont {Chikkadi}}, \bibinfo
  {author} {\bibfnamefont {A.~L.}\ \bibnamefont {Sellerio}}, \bibinfo {author}
  {\bibfnamefont {S.}~\bibnamefont {Zapperi}},\ and\ \bibinfo {author}
  {\bibfnamefont {P.}~\bibnamefont {Schall}},\ }\bibfield  {title} {\bibinfo
  {title} {Direct observation of percolation in the yielding transition of
  colloidal glasses},\ }\href@noop {} {\bibfield  {journal} {\bibinfo
  {journal} {Physical review letters}\ }\textbf {\bibinfo {volume} {118}},\
  \bibinfo {pages} {148001} (\bibinfo {year} {2017})}\BibitemShut {NoStop}%
\bibitem [{\citenamefont {Schall}\ \emph {et~al.}(2007)\citenamefont {Schall},
  \citenamefont {Weitz},\ and\ \citenamefont {Spaepen}}]{schall2007structural}%
  \BibitemOpen
  \bibfield  {author} {\bibinfo {author} {\bibfnamefont {P.}~\bibnamefont
  {Schall}}, \bibinfo {author} {\bibfnamefont {D.~A.}\ \bibnamefont {Weitz}},\
  and\ \bibinfo {author} {\bibfnamefont {F.}~\bibnamefont {Spaepen}},\
  }\bibfield  {title} {\bibinfo {title} {Structural rearrangements that govern
  flow in colloidal glasses},\ }\href@noop {} {\bibfield  {journal} {\bibinfo
  {journal} {Science}\ }\textbf {\bibinfo {volume} {318}},\ \bibinfo {pages}
  {1895} (\bibinfo {year} {2007})}\BibitemShut {NoStop}%
\bibitem [{\citenamefont {Mayr}(2009)}]{mayr2009relaxation}%
  \BibitemOpen
  \bibfield  {author} {\bibinfo {author} {\bibfnamefont {S.}~\bibnamefont
  {Mayr}},\ }\bibfield  {title} {\bibinfo {title} {Relaxation kinetics and
  mechanical stability of metallic glasses and supercooled melts},\ }\href@noop
  {} {\bibfield  {journal} {\bibinfo  {journal} {Physical Review B}\ }\textbf
  {\bibinfo {volume} {79}},\ \bibinfo {pages} {060201} (\bibinfo {year}
  {2009})}\BibitemShut {NoStop}%
\bibitem [{\citenamefont {Zhang}\ \emph {et~al.}(2008)\citenamefont {Zhang},
  \citenamefont {Zhou}, \citenamefont {Lin}, \citenamefont {Chen},\ and\
  \citenamefont {Liaw}}]{zhang2008solid}%
  \BibitemOpen
  \bibfield  {author} {\bibinfo {author} {\bibfnamefont {Y.}~\bibnamefont
  {Zhang}}, \bibinfo {author} {\bibfnamefont {Y.~J.}\ \bibnamefont {Zhou}},
  \bibinfo {author} {\bibfnamefont {J.~P.}\ \bibnamefont {Lin}}, \bibinfo
  {author} {\bibfnamefont {G.~L.}\ \bibnamefont {Chen}},\ and\ \bibinfo
  {author} {\bibfnamefont {P.~K.}\ \bibnamefont {Liaw}},\ }\bibfield  {title}
  {\bibinfo {title} {Solid-solution phase formation rules for multi-component
  alloys},\ }\href@noop {} {\bibfield  {journal} {\bibinfo  {journal} {Advanced
  engineering materials}\ }\textbf {\bibinfo {volume} {10}},\ \bibinfo {pages}
  {534} (\bibinfo {year} {2008})}\BibitemShut {NoStop}%
\bibitem [{\citenamefont {Richard}\ \emph {et~al.}(2020)\citenamefont
  {Richard}, \citenamefont {Ozawa}, \citenamefont {Patinet}, \citenamefont
  {Stanifer}, \citenamefont {Shang}, \citenamefont {Ridout}, \citenamefont
  {Xu}, \citenamefont {Zhang}, \citenamefont {Morse}, \citenamefont {Barrat}
  \emph {et~al.}}]{richard2020predicting}%
  \BibitemOpen
  \bibfield  {author} {\bibinfo {author} {\bibfnamefont {D.}~\bibnamefont
  {Richard}}, \bibinfo {author} {\bibfnamefont {M.}~\bibnamefont {Ozawa}},
  \bibinfo {author} {\bibfnamefont {S.}~\bibnamefont {Patinet}}, \bibinfo
  {author} {\bibfnamefont {E.}~\bibnamefont {Stanifer}}, \bibinfo {author}
  {\bibfnamefont {B.}~\bibnamefont {Shang}}, \bibinfo {author} {\bibfnamefont
  {S.}~\bibnamefont {Ridout}}, \bibinfo {author} {\bibfnamefont
  {B.}~\bibnamefont {Xu}}, \bibinfo {author} {\bibfnamefont {G.}~\bibnamefont
  {Zhang}}, \bibinfo {author} {\bibfnamefont {P.}~\bibnamefont {Morse}},
  \bibinfo {author} {\bibfnamefont {J.-L.}\ \bibnamefont {Barrat}}, \emph
  {et~al.},\ }\bibfield  {title} {\bibinfo {title} {Predicting plasticity in
  disordered solids from structural indicators},\ }\href@noop {} {\bibfield
  {journal} {\bibinfo  {journal} {Physical Review Materials}\ }\textbf
  {\bibinfo {volume} {4}},\ \bibinfo {pages} {113609} (\bibinfo {year}
  {2020})}\BibitemShut {NoStop}%
\bibitem [{\citenamefont {Cubuk}\ \emph {et~al.}(2015)\citenamefont {Cubuk},
  \citenamefont {Schoenholz}, \citenamefont {Rieser}, \citenamefont {Malone},
  \citenamefont {Rottler}, \citenamefont {Durian}, \citenamefont {Kaxiras},\
  and\ \citenamefont {Liu}}]{cubuk2015identifying}%
  \BibitemOpen
  \bibfield  {author} {\bibinfo {author} {\bibfnamefont {E.~D.}\ \bibnamefont
  {Cubuk}}, \bibinfo {author} {\bibfnamefont {S.~S.}\ \bibnamefont
  {Schoenholz}}, \bibinfo {author} {\bibfnamefont {J.~M.}\ \bibnamefont
  {Rieser}}, \bibinfo {author} {\bibfnamefont {B.~D.}\ \bibnamefont {Malone}},
  \bibinfo {author} {\bibfnamefont {J.}~\bibnamefont {Rottler}}, \bibinfo
  {author} {\bibfnamefont {D.~J.}\ \bibnamefont {Durian}}, \bibinfo {author}
  {\bibfnamefont {E.}~\bibnamefont {Kaxiras}},\ and\ \bibinfo {author}
  {\bibfnamefont {A.~J.}\ \bibnamefont {Liu}},\ }\bibfield  {title} {\bibinfo
  {title} {Identifying structural flow defects in disordered solids using
  machine-learning methods},\ }\href@noop {} {\bibfield  {journal} {\bibinfo
  {journal} {Physical review letters}\ }\textbf {\bibinfo {volume} {114}},\
  \bibinfo {pages} {108001} (\bibinfo {year} {2015})}\BibitemShut {NoStop}%
\bibitem [{\citenamefont {Fan}\ \emph {et~al.}(2020)\citenamefont {Fan},
  \citenamefont {Ding},\ and\ \citenamefont {Ma}}]{fan2020machine}%
  \BibitemOpen
  \bibfield  {author} {\bibinfo {author} {\bibfnamefont {Z.}~\bibnamefont
  {Fan}}, \bibinfo {author} {\bibfnamefont {J.}~\bibnamefont {Ding}},\ and\
  \bibinfo {author} {\bibfnamefont {E.}~\bibnamefont {Ma}},\ }\bibfield
  {title} {\bibinfo {title} {Machine learning bridges local static structure
  with multiple properties in metallic glasses},\ }\href@noop {} {\bibfield
  {journal} {\bibinfo  {journal} {Materials Today}\ }\textbf {\bibinfo {volume}
  {40}},\ \bibinfo {pages} {48} (\bibinfo {year} {2020})}\BibitemShut {NoStop}%
\bibitem [{\citenamefont {Boattini}\ \emph {et~al.}(2020)\citenamefont
  {Boattini}, \citenamefont {Mar{\'\i}n-Aguilar}, \citenamefont {Mitra},
  \citenamefont {Foffi}, \citenamefont {Smallenburg},\ and\ \citenamefont
  {Filion}}]{boattini2020autonomously}%
  \BibitemOpen
  \bibfield  {author} {\bibinfo {author} {\bibfnamefont {E.}~\bibnamefont
  {Boattini}}, \bibinfo {author} {\bibfnamefont {S.}~\bibnamefont
  {Mar{\'\i}n-Aguilar}}, \bibinfo {author} {\bibfnamefont {S.}~\bibnamefont
  {Mitra}}, \bibinfo {author} {\bibfnamefont {G.}~\bibnamefont {Foffi}},
  \bibinfo {author} {\bibfnamefont {F.}~\bibnamefont {Smallenburg}},\ and\
  \bibinfo {author} {\bibfnamefont {L.}~\bibnamefont {Filion}},\ }\bibfield
  {title} {\bibinfo {title} {Autonomously revealing hidden local structures in
  supercooled liquids},\ }\href@noop {} {\bibfield  {journal} {\bibinfo
  {journal} {Nature communications}\ }\textbf {\bibinfo {volume} {11}},\
  \bibinfo {pages} {1} (\bibinfo {year} {2020})}\BibitemShut {NoStop}%
\bibitem [{\citenamefont {Ding}\ \emph
  {et~al.}(2014{\natexlab{a}})\citenamefont {Ding}, \citenamefont {Patinet},
  \citenamefont {Falk}, \citenamefont {Cheng},\ and\ \citenamefont
  {Ma}}]{ding2014soft}%
  \BibitemOpen
  \bibfield  {author} {\bibinfo {author} {\bibfnamefont {J.}~\bibnamefont
  {Ding}}, \bibinfo {author} {\bibfnamefont {S.}~\bibnamefont {Patinet}},
  \bibinfo {author} {\bibfnamefont {M.~L.}\ \bibnamefont {Falk}}, \bibinfo
  {author} {\bibfnamefont {Y.}~\bibnamefont {Cheng}},\ and\ \bibinfo {author}
  {\bibfnamefont {E.}~\bibnamefont {Ma}},\ }\bibfield  {title} {\bibinfo
  {title} {Soft spots and their structural signature in a metallic glass},\
  }\href@noop {} {\bibfield  {journal} {\bibinfo  {journal} {Proceedings of the
  National Academy of Sciences}\ }\textbf {\bibinfo {volume} {111}},\ \bibinfo
  {pages} {14052} (\bibinfo {year} {2014}{\natexlab{a}})}\BibitemShut {NoStop}%
\bibitem [{\citenamefont {Ding}\ \emph
  {et~al.}(2014{\natexlab{b}})\citenamefont {Ding}, \citenamefont {Cheng},\
  and\ \citenamefont {Ma}}]{ding2014full}%
  \BibitemOpen
  \bibfield  {author} {\bibinfo {author} {\bibfnamefont {J.}~\bibnamefont
  {Ding}}, \bibinfo {author} {\bibfnamefont {Y.-Q.}\ \bibnamefont {Cheng}},\
  and\ \bibinfo {author} {\bibfnamefont {E.}~\bibnamefont {Ma}},\ }\bibfield
  {title} {\bibinfo {title} {Full icosahedra dominate local order in cu64zr34
  metallic glass and supercooled liquid},\ }\href@noop {} {\bibfield  {journal}
  {\bibinfo  {journal} {Acta materialia}\ }\textbf {\bibinfo {volume} {69}},\
  \bibinfo {pages} {343} (\bibinfo {year} {2014}{\natexlab{b}})}\BibitemShut
  {NoStop}%
\bibitem [{\citenamefont {Ding}\ \emph {et~al.}(2016)\citenamefont {Ding},
  \citenamefont {Cheng}, \citenamefont {Sheng}, \citenamefont {Asta},
  \citenamefont {Ritchie},\ and\ \citenamefont {Ma}}]{ding2016universal}%
  \BibitemOpen
  \bibfield  {author} {\bibinfo {author} {\bibfnamefont {J.}~\bibnamefont
  {Ding}}, \bibinfo {author} {\bibfnamefont {Y.-Q.}\ \bibnamefont {Cheng}},
  \bibinfo {author} {\bibfnamefont {H.}~\bibnamefont {Sheng}}, \bibinfo
  {author} {\bibfnamefont {M.}~\bibnamefont {Asta}}, \bibinfo {author}
  {\bibfnamefont {R.~O.}\ \bibnamefont {Ritchie}},\ and\ \bibinfo {author}
  {\bibfnamefont {E.}~\bibnamefont {Ma}},\ }\bibfield  {title} {\bibinfo
  {title} {Universal structural parameter to quantitatively predict metallic
  glass properties},\ }\href@noop {} {\bibfield  {journal} {\bibinfo  {journal}
  {Nature communications}\ }\textbf {\bibinfo {volume} {7}},\ \bibinfo {pages}
  {1} (\bibinfo {year} {2016})}\BibitemShut {NoStop}%
\bibitem [{\citenamefont {Mizuno}\ \emph {et~al.}(2013)\citenamefont {Mizuno},
  \citenamefont {Mossa},\ and\ \citenamefont {Barrat}}]{mizuno2013measuring}%
  \BibitemOpen
  \bibfield  {author} {\bibinfo {author} {\bibfnamefont {H.}~\bibnamefont
  {Mizuno}}, \bibinfo {author} {\bibfnamefont {S.}~\bibnamefont {Mossa}},\ and\
  \bibinfo {author} {\bibfnamefont {J.-L.}\ \bibnamefont {Barrat}},\ }\bibfield
   {title} {\bibinfo {title} {Measuring spatial distribution of the local
  elastic modulus in glasses},\ }\href@noop {} {\bibfield  {journal} {\bibinfo
  {journal} {Physical Review E}\ }\textbf {\bibinfo {volume} {87}},\ \bibinfo
  {pages} {042306} (\bibinfo {year} {2013})}\BibitemShut {NoStop}%
\bibitem [{\citenamefont {Tsamados}\ \emph {et~al.}(2009)\citenamefont
  {Tsamados}, \citenamefont {Tanguy}, \citenamefont {Goldenberg},\ and\
  \citenamefont {Barrat}}]{tsamados2009local}%
  \BibitemOpen
  \bibfield  {author} {\bibinfo {author} {\bibfnamefont {M.}~\bibnamefont
  {Tsamados}}, \bibinfo {author} {\bibfnamefont {A.}~\bibnamefont {Tanguy}},
  \bibinfo {author} {\bibfnamefont {C.}~\bibnamefont {Goldenberg}},\ and\
  \bibinfo {author} {\bibfnamefont {J.-L.}\ \bibnamefont {Barrat}},\ }\bibfield
   {title} {\bibinfo {title} {Local elasticity map and plasticity in a model
  lennard-jones glass},\ }\href@noop {} {\bibfield  {journal} {\bibinfo
  {journal} {Physical Review E}\ }\textbf {\bibinfo {volume} {80}},\ \bibinfo
  {pages} {026112} (\bibinfo {year} {2009})}\BibitemShut {NoStop}%
\bibitem [{\citenamefont {Ozawa}\ \emph {et~al.}(2018)\citenamefont {Ozawa},
  \citenamefont {Berthier}, \citenamefont {Biroli}, \citenamefont {Rosso},\
  and\ \citenamefont {Tarjus}}]{ozawa2018random}%
  \BibitemOpen
  \bibfield  {author} {\bibinfo {author} {\bibfnamefont {M.}~\bibnamefont
  {Ozawa}}, \bibinfo {author} {\bibfnamefont {L.}~\bibnamefont {Berthier}},
  \bibinfo {author} {\bibfnamefont {G.}~\bibnamefont {Biroli}}, \bibinfo
  {author} {\bibfnamefont {A.}~\bibnamefont {Rosso}},\ and\ \bibinfo {author}
  {\bibfnamefont {G.}~\bibnamefont {Tarjus}},\ }\bibfield  {title} {\bibinfo
  {title} {Random critical point separates brittle and ductile yielding
  transitions in amorphous materials},\ }\href@noop {} {\bibfield  {journal}
  {\bibinfo  {journal} {Proceedings of the National Academy of Sciences}\
  }\textbf {\bibinfo {volume} {115}},\ \bibinfo {pages} {6656} (\bibinfo {year}
  {2018})}\BibitemShut {NoStop}%
\bibitem [{\citenamefont {Chen}\ and\ \citenamefont
  {Schweizer}(2011)}]{chen2011theory}%
  \BibitemOpen
  \bibfield  {author} {\bibinfo {author} {\bibfnamefont {K.}~\bibnamefont
  {Chen}}\ and\ \bibinfo {author} {\bibfnamefont {K.~S.}\ \bibnamefont
  {Schweizer}},\ }\bibfield  {title} {\bibinfo {title} {Theory of yielding,
  strain softening, and steady plastic flow in polymer glasses under constant
  strain rate deformation},\ }\href@noop {} {\bibfield  {journal} {\bibinfo
  {journal} {Macromolecules}\ }\textbf {\bibinfo {volume} {44}},\ \bibinfo
  {pages} {3988} (\bibinfo {year} {2011})}\BibitemShut {NoStop}%
\bibitem [{\citenamefont {Eshelby}(1959)}]{eshelby1959elastic}%
  \BibitemOpen
  \bibfield  {author} {\bibinfo {author} {\bibfnamefont {J.~D.}\ \bibnamefont
  {Eshelby}},\ }\bibfield  {title} {\bibinfo {title} {The elastic field outside
  an ellipsoidal inclusion},\ }\href@noop {} {\bibfield  {journal} {\bibinfo
  {journal} {Proceedings of the Royal Society of London. Series A. Mathematical
  and Physical Sciences}\ }\textbf {\bibinfo {volume} {252}},\ \bibinfo {pages}
  {561} (\bibinfo {year} {1959})}\BibitemShut {NoStop}%
\bibitem [{\citenamefont {Stauffer}\ and\ \citenamefont
  {Aharony}(2018)}]{stauffer2018introduction}%
  \BibitemOpen
  \bibfield  {author} {\bibinfo {author} {\bibfnamefont {D.}~\bibnamefont
  {Stauffer}}\ and\ \bibinfo {author} {\bibfnamefont {A.}~\bibnamefont
  {Aharony}},\ }\href@noop {} {\emph {\bibinfo {title} {Introduction to
  percolation theory}}}\ (\bibinfo  {publisher} {Taylor \& Francis},\ \bibinfo
  {year} {2018})\BibitemShut {NoStop}%
\bibitem [{\citenamefont {Hinrichsen}(2000)}]{hinrichsen2000non}%
  \BibitemOpen
  \bibfield  {author} {\bibinfo {author} {\bibfnamefont {H.}~\bibnamefont
  {Hinrichsen}},\ }\bibfield  {title} {\bibinfo {title} {Non-equilibrium
  critical phenomena and phase transitions into absorbing states},\ }\href@noop
  {} {\bibfield  {journal} {\bibinfo  {journal} {Advances in physics}\ }\textbf
  {\bibinfo {volume} {49}},\ \bibinfo {pages} {815} (\bibinfo {year}
  {2000})}\BibitemShut {NoStop}%
\end{thebibliography}%

\end{document}


\title{Yielding in multi-component metallic glasses: Universal signatures of elastic modulus heterogeneities}

\author{Kamran Karimi$^1$}
\author{Mikko J. Alava$^{1,2}$}
\author{Stefanos Papanikolaou$^1$}
\affiliation{%
 $^1$ NOMATEN Centre of Excellence, National Center for Nuclear Research, ul. A. Sołtana 7, 05-400 Swierk/Otwock, Poland\\
 $^{2}$ Aalto University, Department of Applied Physics, PO Box 11000, 00076 Aalto, Espoo, Finland
}%

\maketitle

\section*{Supplementary Materials}
Simple shear tests were performed on the $xy$ plane at a fixed strain rate $\dot\gamma=10^{-5} ~\text{ps}^{-1}$ and temperature $T=300$ K up to a prescribed strain $\gamma \le 0.2$.  
We subsequently applied six deformation modes $\epsilon^{(m)}$, $m=1...6$ in Cartesian directions $xyz$ described by the following global strain tensors 
\begin{eqnarray}
\bm{\epsilon}^{(1)}&=&\epsilon_{xx}\delta_{\alpha x}\delta_{\beta x}, \nonumber \\ 
\bm{\epsilon}^{(2)}&=&\epsilon_{yy}\delta_{\alpha y}\delta_{\beta y}, \nonumber\\
\bm{\epsilon}^{(3)}&=&\epsilon_{zz}\delta_{\alpha z}\delta_{\beta z}, \nonumber\\
\bm{\epsilon}^{(4)}&=&\epsilon_{xy}(\delta_{\alpha x}\delta_{\beta y}+\delta_{\alpha y}\delta_{\beta x}), \nonumber\\
\bm{\epsilon}^{(5)}&=&\epsilon_{yz}(\delta_{\alpha y}\delta_{\beta z}+\delta_{\alpha z}\delta_{\beta y}), \nonumber\\
\bm{\epsilon}^{(6)}&=&\epsilon_{xz}(\delta_{\alpha x}\delta_{\beta z}+\delta_{\alpha z}\delta_{\beta x}).
\end{eqnarray}
Here $\delta_{\alpha\beta}$ is a Kronecker delta with $\alpha$ and $\beta$ denoting Cartesian indices.
The strain magnitudes are set to $\epsilon_{xx}=\epsilon_{yy}=\epsilon_{zz}=2\epsilon_{xy}=2\epsilon_{yxz}=2\epsilon_{xz}=10^{-2}$ in strain units.
We made use of (deformation-induced) stress differences $\Delta\sigma^{(m)}_{i,\alpha\beta}$, $m=1...6$ in atoms $i=1...N$ to evaluate the corresponding elasticity tensor
\begin{eqnarray}
c^i_{\alpha\beta xx}&=&\Delta\sigma^{(1)}_{i,\alpha\beta}/\epsilon_{xx}, \nonumber \\ 
c^i_{\alpha\beta yy}&=&\Delta\sigma^{(2)}_{i,\alpha\beta}/\epsilon_{yy}, \nonumber \\ 
c^i_{\alpha\beta zz}&=&\Delta\sigma^{(3)}_{i,\alpha\beta}/\epsilon_{zz}, \nonumber \\ 
c^i_{\alpha\beta xy}&=&\Delta\sigma^{(4)}_{i,\alpha\beta}/2\epsilon_{xy}, \nonumber \\ 
c^i_{\alpha\beta yz}&=&\Delta\sigma^{(5)}_{i,\alpha\beta}/2\epsilon_{yz}, \nonumber \\ 
c^i_{\alpha\beta xz}&=&\Delta\sigma^{(6)}_{i,\alpha\beta}/2\epsilon_{xz}.
\end{eqnarray}
It was also checked that the obtained elastic constants are almost insensitive to our choice of the above strain steps (i.e. the so-called \emph{linear} regime).

\begin{figure*}
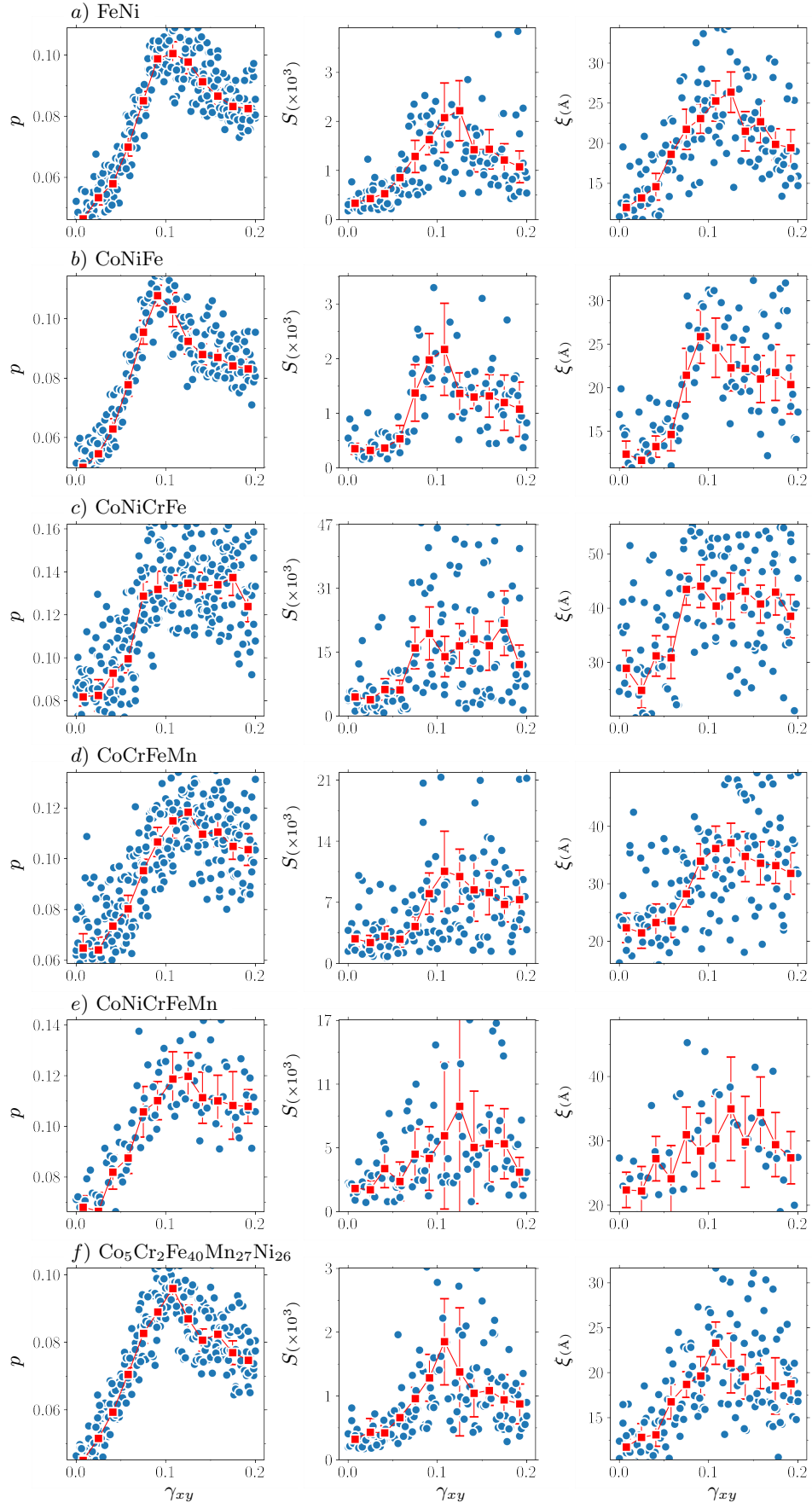

    \foreach \n in {0,...,5}{
    \begin{overpic}[width=0.23\textwidth]{Figs/p_gamma_glass\n.png}
        \LabelFig{12}{87}{\glass{\n}}
        \ifnum\n=5{\Labelxy{50}{-3}{0}{$\gamma_{xy}$}} \fi
        \Labelxy{-8}{46}{90}{{$p$}}
    \end{overpic}
    \begin{overpic}[width=0.23\textwidth]{Figs/smean_gamma_glass\n.png}
        \ifnum\n=5{\Labelxy{50}{-3}{0}{$\gamma_{xy}$}} \fi
        \Labelxy{-8}{46}{90}{{$S\scriptstyle(\times 10^{3})$}}
    \end{overpic}
    \begin{overpic}[width=0.23\textwidth]{Figs/crltnl_gamma_glass\n.png}
        \ifnum\n=5{\Labelxy{50}{-3}{0}{$\gamma_{xy}$}} \fi
        \Labelxy{-8}{46}{90}{{$\xi$\tiny (\r{A})}}
    \end{overpic}
    
    \vspace{7pt}
    }
    \caption{Cluster statistics: probability $p$ ($1$st column), mean size $S$ ($2$nd column), and correlation length $\xi$ ($3$rd column) plotted against applied strain $\gamma_{xy}$ corresponding to the \glzero, \glone, \gltwo, \glthree, \glfour, and \glfive glasses. The (red) curves indicate binning averaged data.}
    \label{fig:statistics_gamma}
\end{figure*}

Statistics of clusters with $\mu=c^i_{xyxy}<0$ associated with \glzero, \glone, \gltwo, \glthree, \glfour, \glfive metallic glasses are illustrated in Fig.~\ref{fig:statistics_gamma} and \ref{fig:statistics_p}.
The fraction of unstable sites $p$ appears to evolve in a non-monotonic fashion, as shown in the leftmost panels of Fig.~\ref{fig:statistics_gamma}(a-f), with a peak value $p_\text{max}$ that almost coincides with that of the (bulk) shear stress $\sigma_{xy}$ (at $\gamma_{xy}\simeq 0.1$).
Overall, the evolution of $p$ with $\gamma_{xy}$ near $p_\text{max}$ features similar trends across all the compositions.
The center and right panels of Fig.~\ref{fig:statistics_gamma} show the mean cluster size $S$ and correlation length $\xi$ as a function of applied strain $\gamma_{xy}$ associated with different compositions.
Both observables tend to grow on approach to failure revealing slight compositional dependence.
This is further demonstrated in the first and third columns of Fig.~\ref{fig:statistics_p}(a-f) with the mean size and correlation length plotted as a function of $p$.
The rescaled data in the second and fourth columns of Fig.~\ref{fig:statistics_p}(a-f) feature a power-law scaling universally described by $S\propto (p_\text{max}-p)^{-\gamma}$ and $\xi\propto (p_{\text{max}}-p)^{-\nu}$ near \pmax.
Here, we note that every panel in the second and fourth columns shows two decades in $S$ and slightly over one decade in $\xi$.
The estimated critical exponents $\gamma$ and $\nu$ are reported in Table~$1$.

\begin{figure*}
    \foreach \n in {0,...,5}{
    \begin{overpic}[width=0.23\textwidth]{Figs/smean_p\n.png}
        \Labelxy{-8}{46}{90}{{$S\scriptstyle(\times 10^{3})$}}
        \LabelFig{15}{86}{\glass{\n}}
        \ifnum\n=5{\Labelxy{50}{-3}{0}{$p$}}\fi
    \end{overpic}
    \begin{overpic}[width=0.23\textwidth]{Figs/smean_p_rescaled\n.png}
        \Labelxy{-8}{46}{90}{{$S$}}
        \ifnum\n=5{\Labelxy{38}{-3}{0}{$1-p/p_\text{max}$}} \fi
        \ifnum\n=0 {
        \begin{tikzpicture}
          \coordinate (a) at (0,0);
            \node[white] at (a) {\tiny.};               %
            \drawSlope{2.4}{2.8}{0.3}{224}{darkred}{$\hspace{-2pt}\gamma$}
		\end{tikzpicture}} \else \fi
    \end{overpic}
    \begin{overpic}[width=0.23\textwidth]{Figs/crltnl_p\n.png}
        \Labelxy{-8}{46}{90}{{$\xi$\tiny (\r{A})}}
        \ifnum\n=5{\Labelxy{50}{-3}{0}{$p$}}\fi
    \end{overpic}
    \begin{overpic}[width=0.23\textwidth]{Figs/crltnl_p_rescaled\n.png}
        \Labelxy{-8}{46}{90}{{$\xi$\tiny (\r{A})}}
        \ifnum\n=5{\Labelxy{38}{-3}{0}{$1-p/p_\text{max}$}}\fi
        \ifnum\n=0 {
        \begin{tikzpicture}
          \coordinate (a) at (0,0);
            \node[white] at (a) {\tiny.};               %
            \drawSlope{1.3}{2.8}{0.3}{231}{darkred}{$\hspace{-2pt}\nu$}
		\end{tikzpicture}} \else \fi
    \end{overpic}
    
    \vspace{10pt}
    }
    \caption{Cluster statistics: mean size $S$ ($1$st and $2$nd columns) and correlation length $\xi$ ($3$rd and $4$th columns) plotted against probability $p$ and $1-p/p_\text{max}$ corresponding to the \glzero, \glone, \gltwo, \glthree, \glfour, and \glfive glasses. The (red) curves indicate binning averaged data. The dashdotted lines denote power laws $S\propto (p_{\text{max}}-p)^{-\gamma}$  and $\xi\propto (p_{\text{max}}-p)^{-\nu}$. The scaling exponents $\gamma$ and $\nu$ are reported in Table~1.}
    \label{fig:statistics_p}
\end{figure*}